\documentclass[12pt]{article}

\usepackage{amsmath,amssymb,graphicx,multirow,xspace}
\usepackage[colorlinks=true,urlcolor=blue,anchorcolor=blue,citecolor=blue,filecolor=blue,linkcolor=blue,menucolor=blue,pagecolor=blue]{hyperref}
\usepackage[compress,numbers]{natbib}
\usepackage{subfigure, placeins}
\usepackage{booktabs}
\usepackage{blindtext, rotating}
\usepackage{afterpage}
\usepackage{enumitem}
\usepackage{ marvosym }
\usepackage{authblk}
\usepackage{verbatim}
\usepackage{soul} 

\usepackage{floatrow}
\newfloatcommand{capbtabbox}{table}[][\FBwidth]

\usepackage[font=footnotesize,labelfont=bf]{caption}

\usepackage{lineno}

\allowdisplaybreaks
\long\def\symbolfootnote[#1]#2{\begingroup%
\def\thefootnote{\fnsymbol{footnote}}\footnote[#1]{#2}\endgroup}

\newcommand{\newc}{\newcommand}
\newc{\gsim}{\lower.7ex\hbox{$\;\stackrel{\textstyle>}{\sim}\;$}}
\newc{\lsim}{\lower.7ex\hbox{$\;\stackrel{\textstyle<}{\sim}\;$}}
\newc{\gev}{\,{\rm GeV}}
\newc{\mev}{\,{\rm MeV}}
\newc{\ev}{\,{\rm eV}}
\newc{\kev}{\,{\rm keV}}
\newc{\tev}{\,{\rm TeV}}

\newc{\mz}{M_Z}
\newc{\mpl}{M_*}
\newc{\mw}{m_{\rm weak}}
\newc{\nr}[1]{N^c_R{}_{#1}}

\def\beq{\begin{equation}}
\def\eeq{\end{equation}}
\newcommand{\bea}{\begin{eqnarray}\begin{aligned}}
\newcommand{\eea}{\end{aligned}\end{eqnarray}}
\def\bitem{\begin{itemize}}
\def\eitem{\end{itemize}}
\newc{\ie}{{\it i.e.}}          \newc{\etal}{{\it et al.}}
\newc{\eg}{{\it e.g.}}          \newc{\etc}{{\it etc.}}
\newc{\cf}{{\it c.f.}}

 \numberwithin{equation}{section}

\newcommand\fverb{\setbox\fverbbox=\hbox\bgroup\verb}
\newcommand\fverbdo{\egroup\medskip\noindent%
            \fbox{\unhbox\fverbbox}\ }
\newcommand\fverbit{\egroup\item[\fbox{\unhbox\fverbbox}]}
\newbox\fverbbox

\usepackage[usenames,dvipsnames]{xcolor}

\begin{document}

\author[1,2]{Simon Knapen\thanks{smknapen@lbl.gov}}
\author[3,4]{Diego Redigolo\thanks{dredigol@lpthe.jussieu.fr}}
\author[5]{David Shih\thanks{dshih@physics.rutgers.edu}}
\small
\affil[1]{\small{Berkeley Center for Theoretical Physics,
University of California, Berkeley, CA 94720}
}
\affil[2]{ Theoretical Physics Group, Lawrence Berkeley National Laboratory, Berkeley, CA 94720
}
\affil[3]{Sorbonne Universit\'es, UPMC Univ Paris 06, UMR 7589, LPTHE, F-75005, Paris, France
}
\affil[4]{CNRS, UMR 7589, LPTHE, F-75005, Paris, France
}
\affil[5]{ \small{New High Energy Theory Center, Rutgers University, Piscataway,
  NJ 08854}}

\title{General Gauge Mediation at the Weak Scale}

\maketitle

\begin{abstract}

We completely characterize General Gauge Mediation (GGM) at the weak scale by solving all IR constraints over the full parameter space.  This is made possible through a combination of numerical and analytical methods, based on a set of algebraic relations among the IR soft masses derived  from the GGM boundary conditions in the UV. We show how tensions between just a few constraints determine the boundaries of the parameter space: electroweak symmetry breaking (EWSB), the Higgs mass, slepton tachyons, and left-handed stop/sbottom tachyons. While these constraints allow the left-handed squarks to be arbitrarily light, they place strong lower bounds on all of the right-handed squarks. Meanwhile, light EW superpartners are generic  throughout much of the parameter space.  This is especially the case at lower messenger scales, where a positive threshold correction to $m_h$ coming from light Higgsinos and winos is essential in order to satisfy the Higgs mass constraint.

\end{abstract}

\newpage

\tableofcontents

\section{Introduction}\label{Intro}

The recent discovery of a Higgs boson near 125 GeV \cite{Aad:2012tfa,Chatrchyan:2012ufa} has important and far-reaching implications for supersymmetry. In minimal implementations of SUSY (i.e.\ the MSSM),  the stops must now either be very heavy ($\gtrsim 10$ TeV) or have a large trilinear coupling to the Higgs, a so-called `$A$-term' \cite{Hall:2011aa, Heinemeyer:2011aa, Arbey:2011ab, Arbey:2011aa, Draper:2011aa, Carena:2011aa, Cao:2012fz,  Christensen:2012ei, Brummer:2012ns}. Although the heavy stop scenario is trivial to achieve, it is less interesting from both the experimental and the theoretical point of view.  Meanwhile the large $A$-term scenario allows for stops to be observed at the LHC, and it presents interesting challenges for model building.

While many ideas have been explored on how to generate large, multi-TeV $A$-terms from integrating out the messengers of SUSY-breaking  \cite{Evans:2011bea,Evans:2012hg,Kang:2012ra,Craig:2012xp,Abdullah:2012tq,Kim:2012vz,Byakti:2013ti,Craig:2013wga,Evans:2013kxa,Calibbi:2013mka,Jelinski:2013kta,Galon:2013jba,Fischler:2013tva,Knapen:2013zla,Ding:2013pya,Calibbi:2014yha,Basirnia:2015vga,Jelinski:2015gsa,Jelinski:2015voa}, perhaps the simplest mechanism comes from the MSSM itself -- radiatively generating $A$-terms through the MSSM RGEs. This is especially necessary in the context of gauge mediated supersymmetry breaking (GMSB). Indeed, while GMSB naturally solves the SUSY flavor problem and remains one of the most well-motivated frameworks for the origin of SUSY breaking at the weak scale (for a review and many original references, see \cite{Giudice:1998bp}), it predicts that the $A$-terms are essentially zero at the messenger scale. 
 
In this paper, we will perform a systematic and thorough investigation of GMSB in the presence of the $m_h=125$~GeV Higgs constraint.  For this purpose, we will employ the framework of ``General Gauge Mediation" (GGM) developed in \cite{Meade:2008wd,Buican:2008ws}.
There the model-independent parameter space and predictions of gauge mediation were shown to be:\footnote{A common extension of gauge mediation is to include additional Higgs-messenger couplings in order to generate $\mu$ and $B_\mu$, see \cite{Komargodski:2008ax,Craig:2013wga} for a discussion in the context of GGM. This may also generate $A$-terms and modify the boundary conditions for $m_{H_u}^2$ and $m_{H_d}^2$. Such models are beyond the scope of this work; see section \ref{subsec:futuredirections} for further comments.}
\begin{itemize}
\item Flavor universality
\item Negligible $A$-terms and $B_\mu$
\item $\mu$ is ``set by hand" 
\item The sfermion soft masses obey the following relations
\bea
& m_{H_u}^2=  m_{H_d}^2= m_{L}^2\\
& m_Q^2-2m_U^2+m_D^2-m_L^2+m_E^2 =0 \\
& 2m_Q^2-m_U^2-m_D^2-2m_L^2+m_E^2=0
\label{sumrules}
\eea
\end{itemize}
All of these conditions hold at the messenger scale $M_{mess}$ and are generally modified by the RG-running to the weak scale.   They allow for seven independent UV parameters that span the full parameter space plus  $M_{mess}$ itself, which sets the length of the RG-flow. A  convenient choice of parameters is\footnote{In this paper, we assume real gaugino masses and $\mu$ to avoid problems with CP violation, but allow for both positive and negative values for all the soft masses (including $\mu$) in (\ref{UVparameterspace}). We also assume messenger parity in the hidden sector so that $U(1)_Y$ $D$-tadpoles are zero in the UV \cite{Meade:2008wd}.}
\begin{equation}
 M_1,\; M_2,\; M_3,\;  m_{Q}^2,\;  m_{U}^2,\;  m_{L}^2,\; \text{ and } \mu \label{UVparameterspace}
\end{equation}
It was shown in \cite{Buican:2008ws} that the full GGM parameter space can be realized in terms of weakly coupled messenger models. It was further shown in \cite{Draper:2011aa} that if one starts with  zero $A$-terms in the UV, then very large gluino masses and high messenger scales are required to generate large $A$-terms at the weak scale. Our goal in this paper is to build on these works, by exploring how the full set of constraints in the UV (the GGM boundary conditions) and the IR (the Higgs mass, EWSB and a  tachyon-free  spectrum) impact the allowed parameter space.  In a companion paper \cite{GGMcollider} we will study the corresponding LHC phenomenology.
  
While the GGM parameter space (\ref{UVparameterspace}) is a huge reduction in complexity compared to the full 100+ parameters of the MSSM soft SUSY-breaking Lagrangian, it is still challenging to survey it fully. The main reason is that the GGM boundary conditions that lead to (\ref{UVparameterspace}) are defined in the UV at $M_{mess}$ while all of the other constraints are applied in the IR at the weak scale. 
Even after RG evolving from $M_{mess}$ down to the weak scale and imposing the EWSB and Higgs mass conditions, a four dimensional parameter space remains. This is further subject to the requirement of a viable (i.e.\ non-tachyonic) spectrum. Previous attempts have dealt with this challenge primarily by taking various 2D slices of the UV parameter space  \cite{Abel:2009ve,Abel:2010vba,Grajek:2013ola,Rajaraman:2009ga,Carpenter:2008he}. Aside from introducing artificial relations among the parameters, this is also suboptimal because scanning the GGM parameter space in terms of the UV parameters is in general quite inefficient. For instance, the IR constraints might not be automatically satisfied at a generic point in the UV parameter space, or the UV parameters might map to uninteresting IR parameters, e.g.\ where some sfermions are extremely heavy and out of reach of the LHC.

A key idea of this paper is to work directly in terms of an equivalent set of IR soft parameters defined at the weak scale:
\begin{equation}
 M_1,\; M_2,\; A_t,\;  m_{Q_3}^2,\;  m_{U_3}^2,\;  m_{L_3}^2,\; \text{ and } \mu \label{parameterspace}
\end{equation}
In order to efficiently map UV to IR parameters, we make use of a ``transfer matrix" approach to the MSSM RGEs: for fixed $\tan\beta$, $M_{mess}$ and $M_{S}=\sqrt{m_{Q_3}m_{U_3}}$, we integrate the RGEs once and for all and encode the result as the coefficients of a (bi)linear transformation between UV and IR parameters. This approach is quite common in high-scale mediation scenarios, but less so in gauge mediation scenarios. Via the transfer matrix and the GGM boundary conditions, all other IR parameters are determined in terms of those in (\ref{parameterspace}) by a set of algebraic relations. (A subset of these relations -- those that are one-loop RG invariants -- was previously presented and studied in detail in \cite{Carena:2010gr,Carena:2010wv}.) 
Using these IR relations to reduce the MSSM soft masses to (\ref{parameterspace}) will streamline the task of scanning over the parameter space, elucidate the phenomenology of GGM, and clarify the interplay between the various IR constraints. 

We will take two complementary approaches to exploring the GGM parameter space at the weak scale.  Our first method is to perform a high-resolution numerical scan on the parameter space in (\ref{parameterspace}). Since the RGEs depend on $M_1$ only through the small hypercharge coupling, $M_1$ plays very little role in the analysis, and so we set $M_1=1$~TeV throughout.  We explore the role of $M_{mess}$ by defining three benchmark scenarios with ``low", ``medium" and ``high" messenger scales, $M_{mess}=10^{7}$, $10^{11}$ and $10^{15}$~GeV respectively. (Messenger scales higher than $10^{15}$~GeV are not considered because gravity-mediated effects are expected to become important, spoiling the flavor-universal GGM boundary conditions.) Finally, we choose $(m_{Q_3}^2,m_{U_3}^2,M_2)$ to scan finely over. For each choice of these parameters, we use the Higgs mass and EWSB  conditions to eliminate  $A_t$, $m_{L_3}^2$ and $\mu$. We use SoftSUSY \cite{Allanach:2001kg} to take into account all relevant IR threshold corrections. For each point in the stop mass plane, the allowed parameter space is an interval (or collection of intervals) in $M_2$. 

We will also study the GGM parameter space analytically in a simplified approximation, in order to gain deeper insights. First, we will neglect all of the threshold corrections to the EWSB equations and truncate them to tree-level. As we will see, this approximation is surprisingly effective. Second, we will greatly simplify the IR relations by using the one-loop RGEs and by neglecting contributions from hypercharge and the bottom and tau Yukawas. For example, the IR relation for the right-handed slepton mass becomes:
\bea
& m_{E_3}^2 \approx 2m_{L_3}^2+{1\over2}\mu^2 +  {3\over2}(m_{U_3}^2-m_{Q_3}^2)\label{IRrelsfintro} 
 \eea 
A complete list of simplified IR relations and their derivation is given in section \ref{subsec:IRrelations}. They are a central result of this paper, and they will prove to be quite powerful. Together with the tree-level EWSB conditions and an accurate Higgs mass calculation via SoftSUSY, we find that we can understand nearly all of the features of the GGM parameter space in this approximate analytical approach. 

As we will see, the IR relations imply certain orderings of the soft masses. For example, we will show that the first and second generation $Q$ and $U$ squarks are always heavier than their third generation counterparts, and that the $D$ squarks are always heavier than the lightest stop. For the other sparticles, the ordering generally depends on where we are in the parameter space. Most importantly,  if $m_{Q_3}<m_{U_3}$, then according to  (\ref{IRrelsfintro}), left-handed sleptons are always lighter than right-handed sleptons. Meanwhile for $m_{Q_3}>m_{U_3}$, right-handed sleptons are always lighter provided $\mu$ is not too large. Based on these orderings, we show that the boundaries of GGM parameter space are solely determined by the Higgs mass, EWSB, slepton tachyons, and left-handed stop/sbottom tachyons. All other potential constraints (such as tachyons from the other scalars) are irrelevant. 

\begin{figure}[t]\centering
\includegraphics[width= 0.48\textwidth]{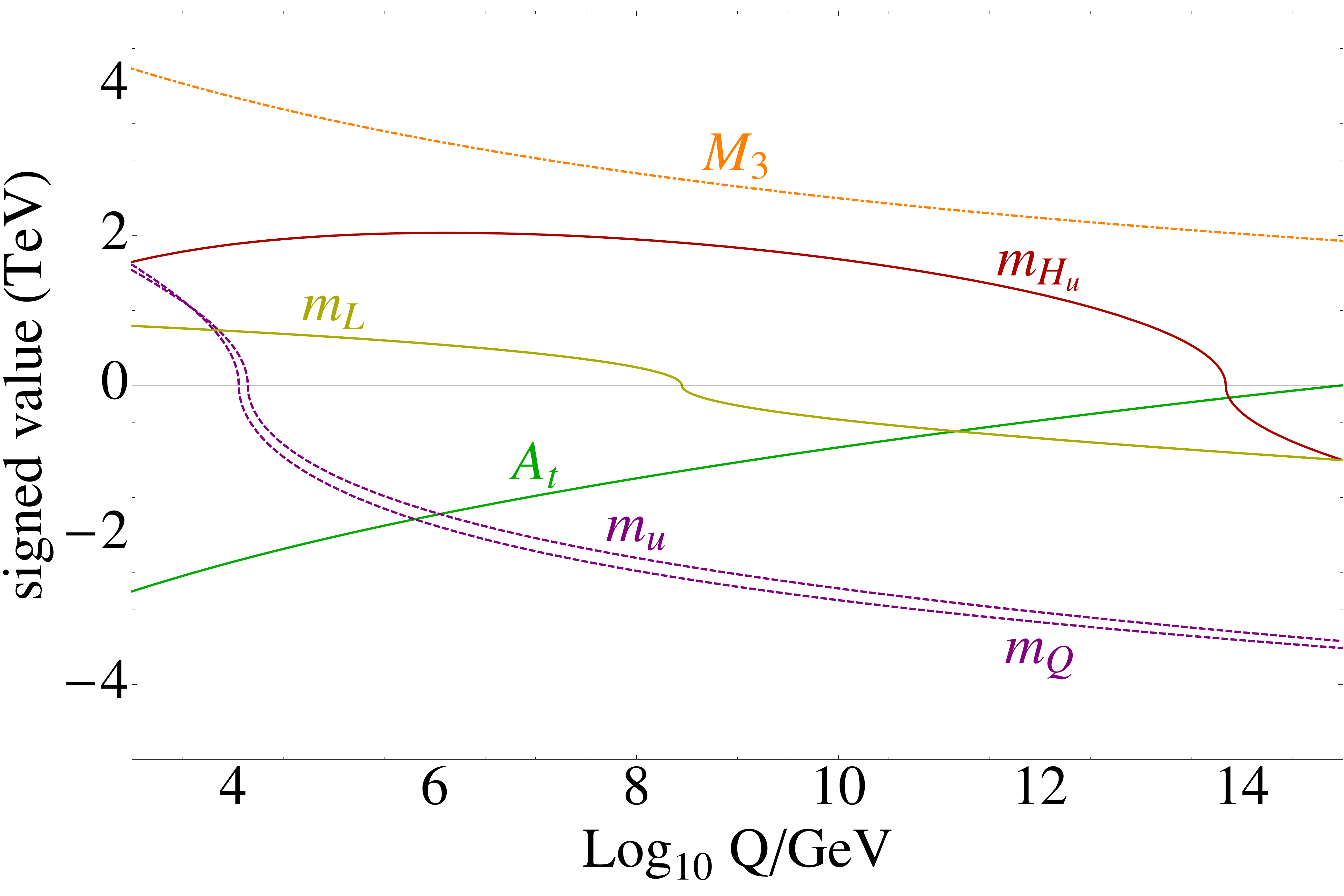}\hfill
\includegraphics[width=0.48\textwidth]{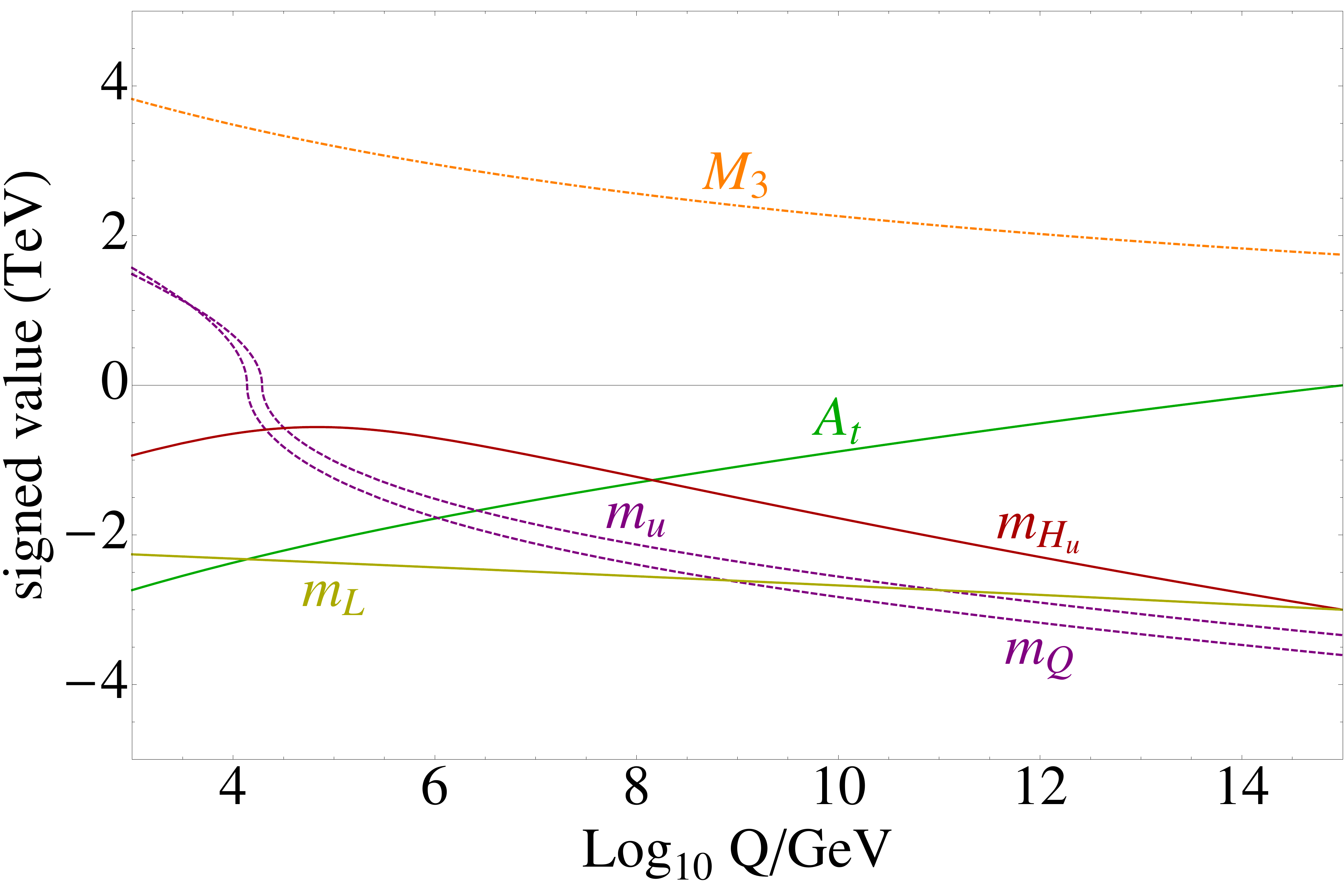}
\caption{Examples of spectra where both $m_{Q_3}$ and  $m_{U_3}$ are small at the weak scale. $m_{H_{u}}$ is driven upward in the RG-running and EWSB is not achieved in the IR (left panel) or the left-handed slepton remains tachyonic in the IR (right panel).\label{RGEplot1}}
\end{figure}

Not surprisingly, the Higgs mass constraint plays an especially important role. The reason is that, as noted above, large radiative $A$-terms with light stops require very heavy gluinos in GGM. Such heavy gluinos have a number of effects on other soft parameters through the RGEs. For example, as was noted in \cite{Draper:2011aa},  the stops must be tachyonic at the messenger scale and over much of the RG. (See also the nice discussion in the earlier work of \cite{Dermisek:2006ey} and its possible implications for fine-tuning.) This is in  tension with EWSB, since negative soft masses for the stops drive $m^2_{H_u}$ \emph{upwards} in the RG-running,
\begin{equation}\label{eq:mhuRGE}
16\pi^2 {d\over dt} m_{H_u}^2 = 6y_t^2(m_{Q_3}^2+m_{U_3}^2) + \dots
\end{equation}
while EWSB requires $m^2_{H_u}<0$ at the weak scale. Of course, the simple way out is to start with sufficiently negative $m^2_{H_u}$ already at the messenger scale. But in models of GMSB, $m_{L_3}^2= m_{H_u}^2$  at the messenger scale, and so left-handed slepton tachyons come into play, ruling out combinations of stop masses and $A$-terms which would otherwise have satisfied the $m_h=125$~GeV constraint.  This logic is further illustrated in fig.~\ref{RGEplot1} for an example point with low stop masses.

\begin{figure}[t]
\includegraphics[width=0.5\textwidth]{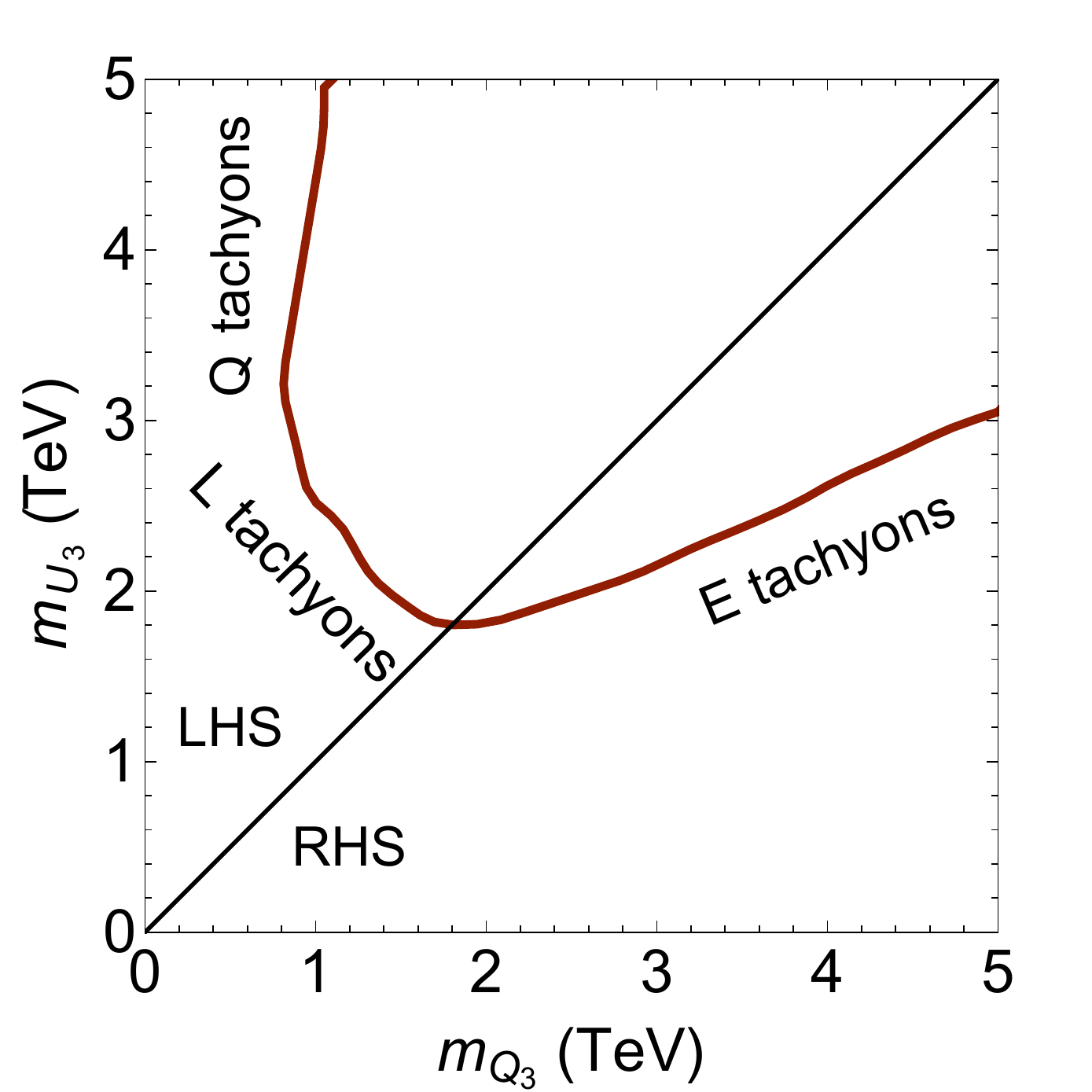}
\caption{The anatomy of the  IR stop mass plane in GGM. The red line schematically indicates the boundary of the viable parameter space; for the actual bounds we refer to fig.~\ref{moneyplane}. The GGM sum rules divide the plane along the diagonal, inducing $L$ tachyons on the left and $E$ tachyons on the right of the diagonal. On the far left, the parameter space is cut off by left-handed stop/sbottom tachyons coming from gluino-induced threshold corrections.\label{cartoon}}
\end{figure}

Given the role of the stop masses in determining not only the Higgs mass, but also the ordering of the slepton masses, the projection of the GGM parameter space into the stop soft-mass plane $(m_{Q_3},\, m_{U_3})$ will prove to be extremely useful throughout this paper. A schematic representation of the stop mass plane is shown in fig.~\ref{cartoon}. We have divided it into two halves along the diagonal, and we will refer to the $m_{Q_3}<m_{U_3}$ ($m_{Q_3}>m_{U_3}$) half as the ``LHS" (``RHS") of the stop mass plane. According to our discussion above, on the LHS (RHS), left-handed (right-handed) slepton tachyons take precedence in determining the boundaries of parameter space. 

These tachyon constraints do not act symmetrically across the diagonal of the stop mass plane. In fact, we will show that the right-handed slepton tachyon constraint leads to a strict lower bound on $m_{U_3}$ of $\gtrsim 1.5$~TeV (and becoming even more stringent with decreasing messenger scale). Because of the IR relations, there are similarly stringent bounds for all of the other right-handed squarks (both up and down-type). Meanwhile, no comparable lower bound on $m_{Q_3}$ exists on the LHS.  Instead, here the boundary arises because a large hierarchy between $m_{Q_3}$ and $M_3$  induces a large, negative threshold correction to the left-handed stop/sbottom mass, driving it tachyonic. The left-handed squark masses of the first/second generation track $m_{Q_3}$, again because of the IR relations. As a result, we find that all three generations of left-handed squarks can be arbitrarily light, despite the constraints on GGM parameter space.

Near the boundaries of parameter space, a convergence of constraints leads to a highly predictive set of spectra with definite implications for the collider phenomenology. In particular, since the boundaries are always determined in part by a sparticle mass going tachyonic, the spectrum there is always characterized by light sparticles. For instance, the $L$ and $E$ boundaries are always accompanied by relatively light left and right handed sleptons respectively, while the $Q$ boundary predicts light left-handed stops/sbottoms. The tension with EWSB generally implies light Higgsinos as well. These are interesting predictions for the boundary of the GGM parameter space and provide additional motivation for LHC searches focused on stops, sbottoms and EW superpartners.

Finally, we will see from both the full numerical scan and the approximate analytic approach that the sign of $\mu$ is an important discrete choice that affects many qualitative features of the GGM spectrum. Requiring that there be no pseudoscalar tachyons correlates  the sign of $\mu$ and the range of $M_2$ throughout GGM parameter space;  in particular, only for $\mu>0$ can one obtain $M_2=0$. The Higgs mass receives a $\sim 2$--3~GeV boost from light charginos and neutralinos in the neighborhood of $M_2= 0$, and this leads to a significant decrease in the required $A_t$ for $\mu>0$ compared to $\mu<0$. We will see how this difference between the two signs of $\mu$  becomes more striking as $M_{mess}$ is decreased and all the various constraints become much stronger. Eventually, the positive chargino/neutralino threshold correction becomes critical to fulfill the Higgs mass constraint and $\mu>0$ with $M_2\approx0$ dominates the parameter space. Light winos are therefore another robust feature of the GGM parameter space with lower $M_{mess}$, with possibly important consequences for the LHC phenomenology.

The remainder of this paper is organized as follows. In section \ref{sec:generalities} we discuss the general features of the GGM parameter space at the weak scale, and in particular the impact of imposing the Higgs mass constraint. We derive the approximate IR relations that result from the GGM boundary conditions in the UV. We use these to identify the relevant constraints (EWSB, Higgs, slepton and left-handed stop/sbottom tachyons) and show how they restrict the parameter space. Section \ref{sec:resultsgeneral} contains the methodology and results of our numerical scan with SoftSUSY. We present results in the stop mass plane, and also along several benchmark slices of parameter space, which serve to further  illustrate the features of the parameter space and the impact of the various constraints. Section \ref{sec:interpretation} is devoted to a detailed semi-analytic study of the parameter space, which we use to validate and elucidate features of the full numerical scan. We conclude in section \ref{sec:conclusions} with a summary of our results, a brief preview of the upcoming paper \cite{GGMcollider}  on GGM collider phenomenology,  and a discussion of future directions. Appendix \ref{appALGO} contains validation plots for the transfer matrix and our numerical scan, while appendix \ref{appChargino/neutralino}  discusses  in more detail the chargino/neutralino threshold correction to the Higgs mass.

\section{GGM at the weak scale}
\label{sec:generalities}

\subsection{IR relations in GGM}
\label{subsec:IRrelations}

As described in the introduction, one of the key ideas of this paper is to work directly in terms of the IR soft parameters:
\beq 
M_1,\; M_2,\; A_t, \; m_{Q_3}^2,\; m_{U_3}^2,\; m_{L_3}^2,\;\mu
\label{IRparameters}
\eeq
We are able to do this efficiently by using a ``transfer matrix" approach to the MSSM RGEs. For fixed  $\tan\beta$, $M_{mess}$, and  $M_S=\sqrt{m_{Q_3} m_{U_3}}$,  we can integrate the MSSM RGEs once and for all and  relate the UV soft parameters to the IR ones using a set of simple {\it algebraic} relations. For instance, for $M_{mess}=10^{15}$~GeV, $M_S=3$~TeV and $\tan\beta=20$ we find\footnote{We will fix $\tan\beta=20$ everywhere in this paper, as this saturates the tree-level contribution to the Higgs mass in the MSSM, without being so large that bottom and tau Yukawa effects cannot be neglected. As long as $\tan\beta$ remains moderately large,  we do not expect our conclusions to change much.}
\bea
&m^2_{Q_3}\approx 0.9\; \hat m^2_{Q}+2.3\; \hat M_3^2+0.3\;\hat M_2^2 + \dots\\
& m^2_{U_3}\approx 0.8\; \hat m^2_{U}+2.0\; \hat M_3^2 + \dots\\
& A_t\approx -0.99\;\hat M_3-0.2\; \hat M_2 + \dots  \label{xferstops}
\eea
where $\dots$ refers to contributions with smaller coefficients, and the hatted (unhatted) quantities are the UV (IR) parameters.
To achieve optimal convergence with the remainder of our algorithm, we extracted these coefficients using the full 2-loop RGEs of SoftSUSY-3.5.1 \cite{Allanach:2001kg}.\footnote{The full set of transfer matrix coefficients used in this paper can be accessed in a accompanying {\tt Mathematica} notebook, which is included in the source of this paper on \url{http://arxiv.org/} .}

Using the transfer matrix, we can algebraically reduce all other IR soft parameters to those in (\ref{IRparameters}). These IR relations are the low-energy versions of the GGM boundary conditions. The full set of IR relations is very complicated and we will not reproduce them here. (For a subset of these relations that are renormalization group invariants, independent of the messenger scale, see  the in-depth discussion in \cite{Carena:2010gr,Carena:2010wv}.) Rather, in this subsection, we will study the IR relations in a simplified approximation that consists of using the one-loop RGEs; neglecting $y_b^2$, $y_\tau^2$ and $g_1^2$ corrections; and imposing EWSB at large $\tan\beta$:
 \bea
& m_Z^2=-2(m^2_{H_u}+|\mu|^2)+\dots\\
& B_\mu\tan\beta = m^2_{H_d}-m_{H_u}^2+\dots \label{ewsb1}
\eea
These simplified IR relations will form the basis of our understanding of the GGM parameter space. Note that they are independent of the Higgs mass constraint; we will come to that in the next subsection. All of the sub-leading corrections (and more) are properly taken into account in a full numerical scan using SoftSUSY, to be described in section \ref{sec:resultsgeneral}. However, as we will see through numerous detailed comparisons with this scan, the approximate treatment introduced here manages to capture most of the qualitative and even quantitative features of the parameter space.

We begin with the IR relations for the sfermion masses:
\bea
&m^2_{Q_{1,2\phantom{,3}}}\approx m^2_{Q_3}+{1\over3}(m_{L_3}^2-m_{H_u}^2)\\
&m^2_{U_{1,2\phantom{,3}}} \approx m^2_{U_3}+{2\over 3}(m_{L_3}^2-m_{H_u}^2)\\
& m_{L_{1,2\phantom{,3}}}^2\approx m_{L_3}^2\\
& m_{D_{1,2,3}}^2 \approx {1\over2}(m_{Q_3}^2+m_{U_3}^2)-{1\over2}m_{H_u}^2\\
& m_{E_{1,2,3}}^2 \approx 2m_{L_3}^2-{1\over2}m_{H_u}^2 +  {3\over2}(m_{U_3}^2-m_{Q_3}^2)
\label{IRrelsf} 
\eea
These relations are satisfied exactly at the messenger scale due to the GGM boundary conditions. In the IR, they are only violated by small effects proportional to $y_b^2$ and $y_\tau^2$. Working in the same approximation, we do not concern ourselves with the small splittings amongst the three generations of sleptons and right-handed sbottoms (see \cite{Calibbi:2014pza} for a discussion of the slepton splitting in GGM). Notice that these relations are independent of the messenger scale and the details of the transfer matrix. Thus they are examples of the renormalization group invariants discussed in \cite{Carena:2010gr,Carena:2010wv}.

After imposing the large $\tan\beta$ EWSB condition $m_{H_u}^2\approx -\mu^2$, we reduce the other sfermion masses to simple combinations of the IR parameters in (\ref{IRparameters}). These IR relations have a number of interesting consequences, which we list here:
\begin{itemize}

\item The 1st/2nd generation $Q$ and $U$ squarks are always heavier than their 3rd generation counterparts. We emphasize that this result is not completely trivial once negative mass-squareds in the UV are allowed (as is the case in GGM), as these could a priori reverse the Yukawa effects in the RGEs that usually drive the third generation squarks lighter.

\item The $D$ squarks are always heavier than the root-mean-squared of the stop masses. 

\item The right-handed sleptons are strictly heavier than the left-handed sleptons, provided that $m_{Q_3}<m_{U_3}$. For $m_{Q_3}>m_{U_3}$, which is lighter depends on   $\mu^2$. 

\end{itemize}

Next we turn to the Higgs sector. Here the IR relations are:\footnote{Note that we are assuming $A_t=B_\mu=0$ in the UV for simplicity. In practice there are generally small, higher-loop contributions to these quantities in GGM. We have checked that none of our results depend sensitively on $A_t$ and $B_\mu$ being literally zero in the UV.}
\bea
& m_{H_d}^2\approx m_{L_3}^2\\
 & m_{H_u}^2 \approx    e\,(\delta M_2+d\, A_t )^2+a\, m_{L_3}^2-m_0^2   \\
&m_A^2\approx  B_\mu \tan\beta \approx  -g\,\delta M_2 \, \mu \tan\beta\label{IRrel1higgs} 
\eea
where we have defined 
\beq
\delta M_2 \equiv M_2+f\, A_t \label{deltaM2def}
\eeq
and
\beq
m_0^2\equiv b\,(m_{Q_3}^2+m_{U_3}^2)-c\, A_t^2 \label{m0def}
\eeq
The first relation in (\ref{IRrel1higgs}) is a consequence of the GGM boundary conditions at the messenger scale and it is only violated by $y_b$ effects. The second and third relations are derived by integrating the MSSM RGEs, dropping subdominant contributions proportional to $g_1^2$. Unlike the previous IR relations, these depend on the messenger scale; see table~\ref{tabsumrulesparam} for benchmark values of the coefficients $a$, $b$, \dots.  In terms of (\ref{IRrel1higgs}), the tree-level EWSB equations (\ref{ewsb1}) become
\bea
& e\,(\delta M_2+d\, A_t)^2 + a\,m_{L_3}^2+\mu^2 \approx m_0^2\\
& - g\,\delta M_2\,\mu\tan\beta \approx m_{L_3}^2+\mu^2
\label{ewsbir}
\eea
From these IR relations, we learn that
\begin{table}[t]
$$
\begin{array}{|c|ccccccccc|}\hline
 M_{mess} & \text{a} & \text{b} & \text{c} & \text{d} & \text{e} & \text{f} & \text{g} &\text{p}  &\text{q}  \\\hline
  10^{15}\text{ GeV} & 0.59 & 0.43 & 0.46 & 0.7 & 0.2 & 0.73 & 0.45 & 1.62 & 0.47 \\
 10^{11}\text{ GeV} & 0.69 & 0.33 & 0.49 & 0.83 & 0.1 & 0.87 & 0.28 & 1.95 & 0.39 \\
 10^7\text{ GeV} & 0.83 & 0.18 & 0.44 & 1.51 & 0.02 & 1.09 & 0.13 & 3.08 & 0.31 \\\hline
\end{array}
$$
\caption{Parameters used in the IR relations (\ref{IRrel1higgs})-(\ref{gluinosoft}) for various values of $M_{mess}$, with $\tan\beta=20$ and $M_S=3$ TeV. \label{tabsumrulesparam} }
\end{table}
\begin{itemize}

\item An important  corollary of the formula for $m_A^2$ in (\ref{IRrel1higgs}) is that the sign of $\mu$ and $\delta M_2$ are correlated. Concretely, if $\mu<0$ ($\mu>0$) we must have $\delta M_2>0$ ($\delta M_2<0$) to avoid pseudoscalar tachyons.

\item In fact, pseudoscalar tachyons are always superseded by positivity of $m_{L_3}^2$ and $\mu^2$, according to the second EWSB condition in (\ref{ewsbir}). 

\item  Also from the second line in (\ref{ewsbir}) it is clear that $\mu=0$ is not an independent constraint, at least in our current approximation, since it always implies $m_{L_3}^2=0$.

\item From (\ref{IRrelsf}) and (\ref{ewsbir}), it follows that
\beq
m_{E_3}^2 < \left(\frac{3}{2}+\frac{2 b}{a}\right) m_{U_3}^2 - \left(\frac{3}{2}-\frac{2 b}{a}\right) m_{Q_3}^2-\frac{2c}{a}A_t^2\,.
\label{mEltmU} \eeq
Since $\frac{2 b}{a}<\frac{3}{2}$ for all messenger scales, we expect that ${E}$ tachyons are always a stronger constraint than $U_3$ tachyons. Ultimately this translates into a strong lower bound on $m_{U_3}$, as we will show in section \ref{sec:resultsgeneral}.

\item The quantity $m_0^2$ defined in (\ref{m0def}) must be positive, otherwise the first EWSB condition in (\ref{ewsbir}) cannot be satisfied with non-tachyonic sleptons. This places an upper bound on the magnitude of the $A$-term allowed at each point in the stop mass plane.

\end{itemize}

Finally, let us comment on the role played by the gluino. The IR gluino mass is given in terms of $A_t$ and $\delta M_2$ by
\begin{equation}
M_3\approx-( p\, A_t+ q\, \delta M_2)\label{gluinosoft}
\end{equation}
where benchmark values of $p$ and $q$ are listed in table~\ref{tabsumrulesparam}.  This equation shows how $M_3$ is linearly related to $A_t$ and $\delta M_2$. The proportionality constant $p$ moreover increases with lowered $M_{mess}$. This reflects the fact that a larger gluino mass is needed to achieve the same $A_t$ for a shorter amount of RG running. As we will see in the following sections, enormous gluino masses are generally required to achieve the large $A$-term scenario with lower messenger scales, and this can result in large gluino-induced threshold corrections to the IR squark masses, as given by equation (34) in \cite{Pierce:1996zz}:
\beq
\delta m_{\tilde q}^2=  \frac{g_3^2}{6\pi^2}m_{\tilde q}^2  \; \left(1 + 3 x+(x-1)^2\log|x-1|-x^2\log x+2 x\log\left[\frac{M_S^2}{m_{\tilde q}^2}\right]\right)\label{bmpzform}
\eeq
Here $m_{\tilde q}$ stands for any of the squark soft masses, and $x\equiv M_3^2/m_{\tilde q}^2$. These threshold corrections are generally negative for the gluino masses of interest (i.e for $M_3\gg M_S$), and will eventually turn the physical squark mass tachyonic. As we will see, this effect is ultimately responsible for the left-most boundary in fig.~\ref{cartoon}.

\subsection{Imposing the Higgs mass constraint}\label{subsec:generalpicture}

Now we will impose the Higgs mass constraint and discuss its implications for GGM. Throughout this work, we will require $m_h=123$~GeV, in order to account conservatively for the theory uncertainty \cite{Allanach:2004rh} in the Higgs mass calculation.
In the MSSM, the Higgs mass is given by the well-known formula
\bea
m^2_h&=m_Z^2 \cos^2( 2\beta) +\frac{3v^2}{4\pi^2}\Bigg(|y_t|^4\log\left(\frac{M_S^2}{m^2_t}\right)+\frac{A_t^2}{M_S^2}\left(|y_t|^2-\frac{A^2_t}{12 M_S^2}\right)\Bigg)+\dots
\label{higgsmass}
\eea
Here the $\dots$ denote important additional corrections from $m_{Q_3}\ne m_{U_3}$, other sparticle thresholds and higher loops. These are accounted for in our analysis using SoftSUSY.

\begin{figure}[t!]\centering
\includegraphics[width=0.5\textwidth]{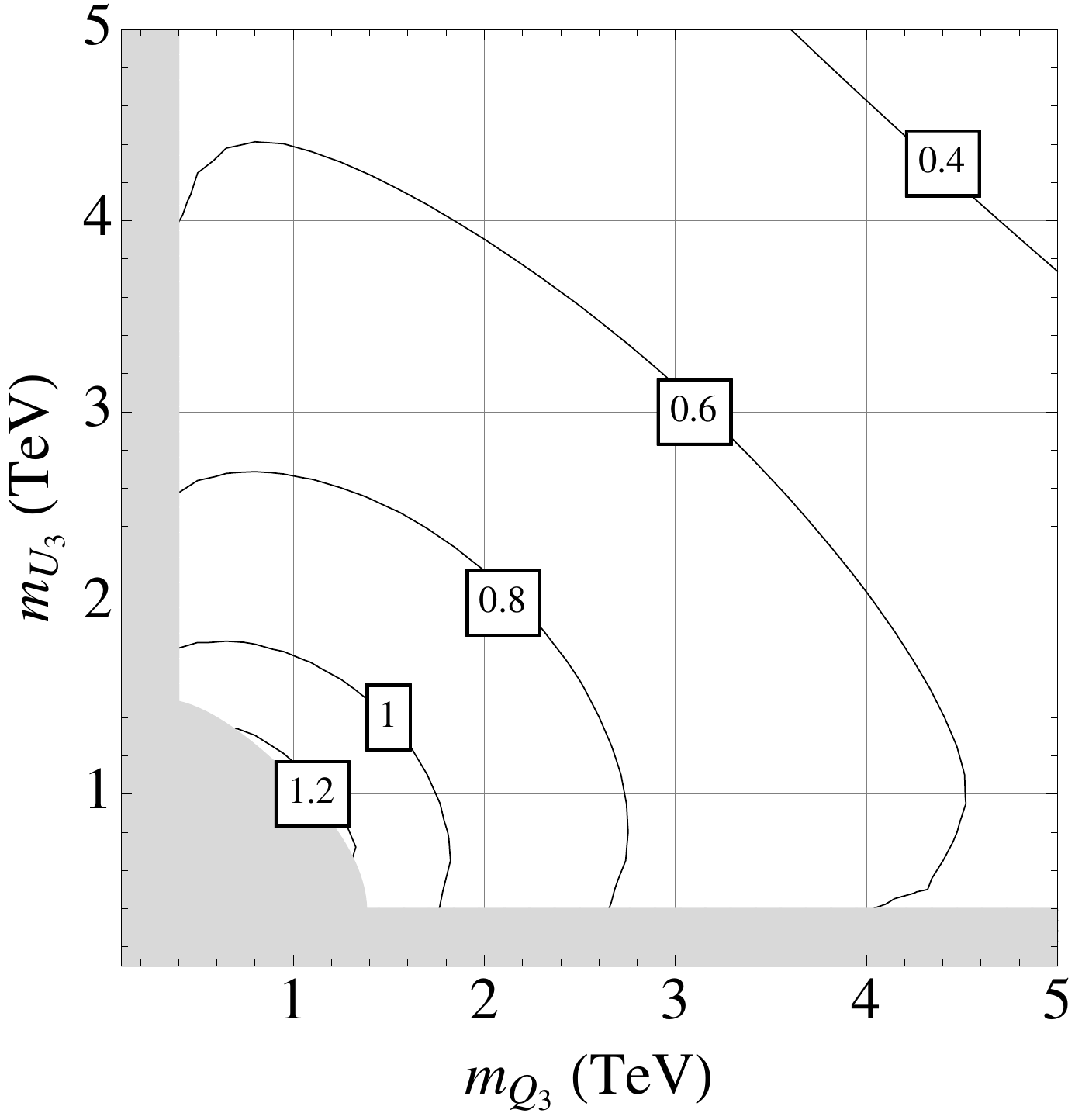}
\caption{Contour plot of the $R_t$ needed  in the MSSM for $m_h=123$~GeV as a function of the stop soft masses. Other soft masses were set to $M_S=\sqrt{m_{Q_3}m_{U_3}}$. The gray regions indicate places where the Higgs mass constraint was impossible to satisfy. In the lower left corner, this is because no $A$-term satisfies the constraint. On the sides, this is because various threshold corrections coming from the very split stops render the output of SoftSUSY unreliable.}\label{mhirgrid}
\end{figure}

The Higgs mass stringently constrains the stop masses and the $A$-terms in the MSSM; for TeV-scale stops, the $A$-terms must generally be multi-TeV. An example of this is given in fig.~\ref{mhirgrid}. Shown here are contours of the ``normalized $A$-term"
\beq\label{Ahat}
R_t\equiv \frac{|A_t|}{\sqrt{m_{Q_3}^2+m_{U_3}^2}}
\eeq
required for $m_h=123$~GeV in SoftSUSY, with all other superpartner masses set to $M_S$.\footnote{We emphasize that this figure is meant to give a general impression and should not be taken literally. The $A$-term required for $m_h=123$~GeV can depend sensitively on the masses of the other superpartners and their contributions to the Higgs mass.}
As we lower the stop masses, the required $R_t$ increases, and for stops below $\sim 1$~TeV, the Higgs mass constraint cannot be satisfied.

In GGM the Higgs mass constraint is even more stringent, since we cannot vary the other parameters of the MSSM arbitrarily. As noted in the previous subsection, for $m_0^2<0$, the first EWSB equation in (\ref{ewsbir}) has no valid solution, since the LHS must be a sum of non-negative quantities. This translates to  the requirement that 
\beq
R_t^2<b/c
\eeq
and from table~\ref{tabsumrulesparam}, we see that $\sqrt{b/c}=1.01,\, 0.85,\, 0.69$ for $M_{mess}=10^{15},\,10^{11},\,10^{7}$~GeV respectively. Comparing with fig.~\ref{mhirgrid}, we see that stop masses that would otherwise be allowed by the Higgs mass constraint are ruled out in GGM by the combination of EWSB and no-tachyon conditions. (Keep in mind that while $m_0^2=0$ furnishes an absolute boundary to the parameter space, there can be even more stringent boundaries  due to tachyon constraints.)

Clearly, the decrease of $\sqrt{b/c}$ with messenger scale amplifies the tension between large $A$-terms and the other constraints.  This will serve to enhance the role of secondary threshold corrections that can increase $m_h$ and allow for smaller $A_t$. As we will see in section \ref{subsec:LElines} (and will discuss further in appendix \ref{appChargino/neutralino}), chief among these is a $\sim 2$~GeV positive threshold correction to $m_h$ coming from light winos and Higgsinos when $M_2$ and $\mu$ are both close to zero. Since the $M_2\approx 0$ region requires $\mu>0$ according to the first bullet point below (\ref{ewsbir}), this will lead to a marked   difference between $\mu<0$ and $\mu>0$ parameter spaces as $M_{mess}$ decreases.

To summarize, we have seen in this section that the IR soft parameters of GGM are related to those in (\ref{IRparameters}) via a set of simple algebraic relations. Some of these IR relations are renormalization group invariants along the lines of \cite{Carena:2010gr,Carena:2010wv}, while others are not. Using these relations, we have shown that the IR soft masses obey certain fixed orderings. In particular, the only soft masses in GGM that can become tachyonic independently of others are $m_{Q_3}^2$, $m_{L_3}^2$ and $m_{E_3}^2$. All other soft masses  are always positive as long as these soft masses are positive. 
Together with the Higgs mass and EWSB constraints, these determine the boundaries of GGM parameter space.

We also showed that important qualitative dividing lines cutting through the parameter space include: the diagonal of the stop mass plane,   the sign of $\mu$, and the (anti-correlated) sign of $\delta M_2$. Using the important variables $m_0^2$ and $R_t$, we saw how decreasing the messenger scale results in increasingly stringent constraints on the parameter space. 
In the following sections, we will confirm this general picture using a high-resolution numerical scan of the GGM parameter space, together with an analytical approach based on the approximate IR relations and tree-level EWSB conditions.

\section{Scanning the GGM parameter space\label{sec:resultsgeneral}}

\subsection{Details of the scan}

\begin{figure}[b!]\centering
\includegraphics[width=1\textwidth]{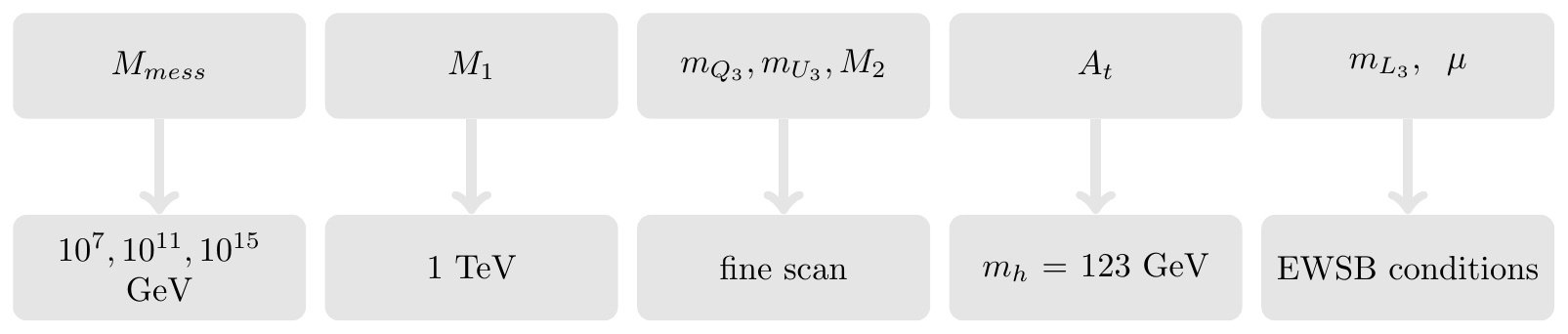}\hfill
\caption{A schematic diagram illustrating the steps in our numerical scan over the GGM parameter space. We trade all UV parameters for the more physical IR parameters in (\ref{IRparameters}) using the transfer matrix and use the EWSB conditions and the Higgs mass constraint to eliminate $A_t,\; m_{L_3}^2$ and $\mu$. We further fix $M_1=1$~TeV, $\tan\beta=20$, scan coarsely over $M_{mess}$, and scan finely over $m_{Q_3},\; m_{U_3}$ and $M_2$.  \label{flowchart}}
\end{figure}

In the introduction, we sketched out the steps in our numerical scan of the GGM parameter space. These steps are summarized in fig.~\ref{flowchart}. Here, we will describe them in more detail. 

\begin{enumerate}

\item We define ``low", ``medium" and ``high" messenger scale benchmarks corresponding to $M_{mess}=10^{7}$, $10^{11}$ and $10^{15}$ GeV respectively.  Moreover, having verified that $\alpha_1$ effects have very little impact on the analysis we set $M_1=1$~TeV throughout this paper.

\item We choose to eliminate $A_t$, $m_{L_3}^2$ and $\mu$ using the Higgs mass and EWSB equations, since the former depends strongly on $A_t$, while the latter are sensitive to $m_{L_3}^2$ and $\mu$. An additional benefit of this choice is that $m_{L_3}^2$ appears linearly in the EWSB equations (\ref{ewsbir}).

\item This leaves $m_{Q_3}$, $m_{U_3}$ and $M_2$ as independent parameters. As described in the introduction, a convenient way to view this remaining parameter space is that for every point in the stop mass plane, all soft parameters are functions defined on an interval or collection of intervals in $M_2$. The $A^0$ tachyon condition cuts the $M_2$ interval into two disconnected pieces, one for each sign of $\mu$. Both pieces are further bounded by requiring the absence of slepton tachyons and by the $\mu^2>0$ conditions. These features are illustrated in fig.~\ref{intervalexample} for an example point in the stop mass plane. 

\item Finally, we perform a high-resolution three-dimensional scan over $(m_{Q_3}, m_{U_3}, M_2)$. Near the boundary of the parameter space the resolution of the scan is further increased, such that this important region is sampled as accurately as possible. The end result is a complete grid of valid spectra spanning the GGM parameter space and satisfying the Higgs mass and all other IR constraints.  Appendix \ref{appALGO} contains several validation plots which demonstrate the convergence of our algorithm.

 \begin{figure}[t]
\centering
\includegraphics[width=0.85\textwidth]{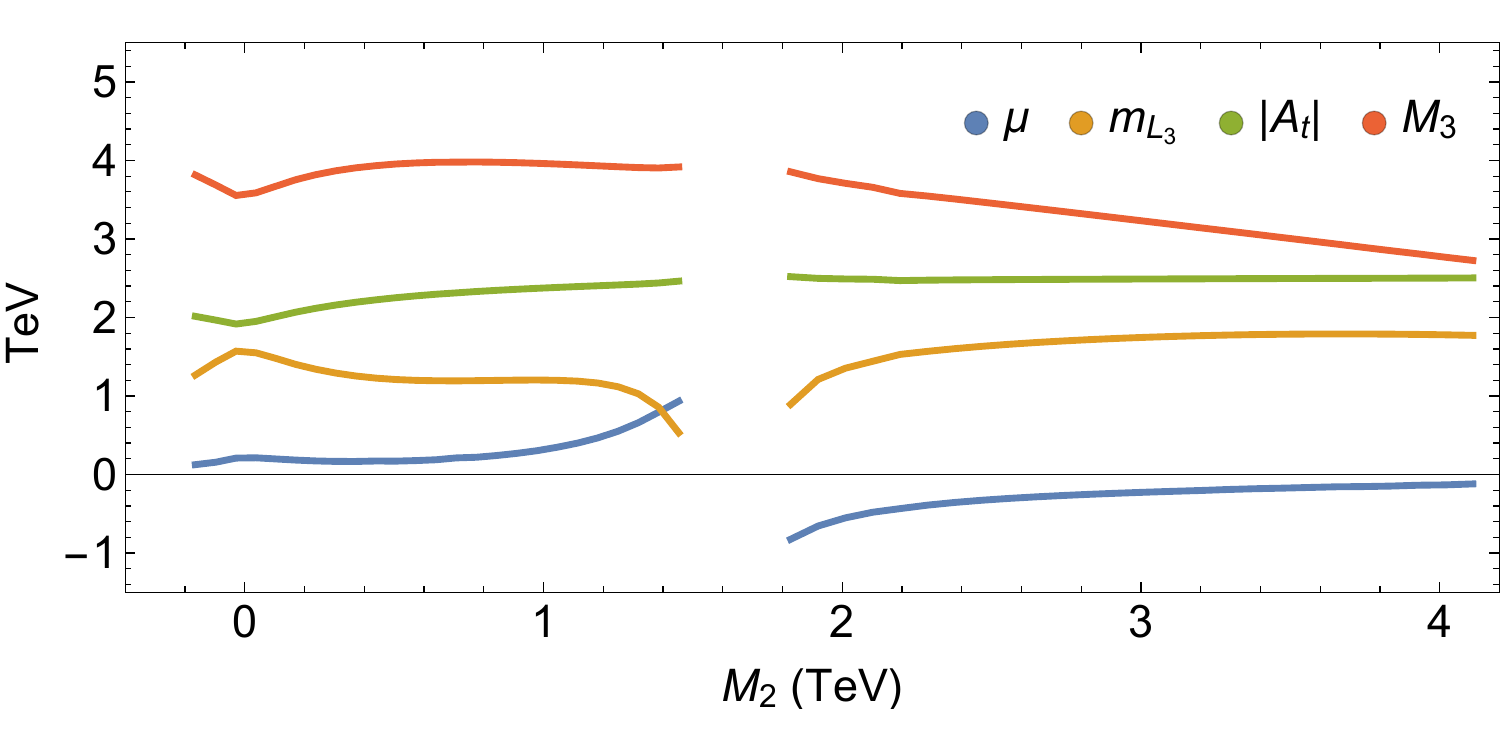}
\caption{Example of the behavior of various soft parameters over the $M_2$ interval. Here $m_{Q_3}=1$ TeV, $m_{U_3}=3$ TeV and $M_{mess}=10^{15}$ GeV. The feature around $M_2\approx 0$ is due to threshold correction to $m_h$ and will be discussed in detail in the next sections.\label{intervalexample} }
\end{figure}

 \end{enumerate}

In practice, step 2 above is the most challenging part of the analysis, because of the complicated threshold corrections that must be taken into account.
Because we use SoftSUSY to implement all the threshold corrections and  RGEs, there is a particular order in which we must solve these constraints. For a given input $m_{H_u}^2$ and $m_{H_d}^2$ at the messenger scale, SoftSUSY imposes the EWSB conditions and returns $\mu$ and $B_\mu$ at the messenger scale. It also computes $m_h$. Thus for each $(m_{Q_3},m_{U_3},M_2)$, $\mu$ is given automatically by SoftSUSY, but we must perform an auxiliary 2D scan over $A_t$ and $m_{L_3}^2$ and numerically solve the $B_\mu({\rm UV})=0$  and $m_h=123$ GeV constraints. 
In principle, this adds two extra dimensions to our scan. A brute force, flat scan over $(A_t,m_{L_3}^2)$ proves to be computationally unfeasible. Instead, we use an iterative method where we sample a few points in the neighborhood of a seed guess, linearly interpolate in $B_\mu({\rm UV})$ and $m_h$ to establish a new seed, and repeat. Typically this converges quickly, after just a few steps, on an extremely accurate solution. Of course, for this to work, it is crucial to obtain an accurate initial seed for $(A_t,m_{L_3}^2)$. We accomplish this by extrapolating from a neighboring point in the $(m_{Q_3},m_{U_3},M_2)$ parameter space.

\subsection{Results: a ``birds-eye view"}
\label{sec:results}

In the rest of this section, we will exhibit the results of the scan outlined above. The primary focus here will be on describing its features;  a fuller analytic understanding in terms of the IR relations (\ref{IRrelsf}) and (\ref{IRrel1higgs}) and the tree-level EWSB  equations (\ref{ewsbir}) will follow in section \ref{sec:interpretation}.

\begin{figure}[t!]\centering
\includegraphics[width=.48\textwidth]{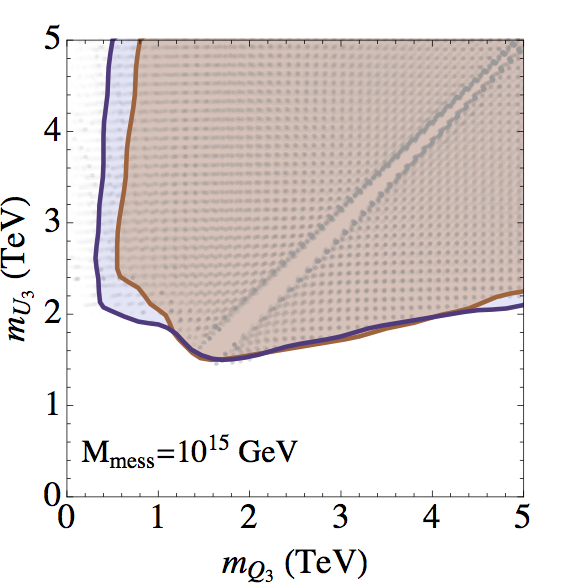}
\hfill
\includegraphics[width=.48\textwidth]{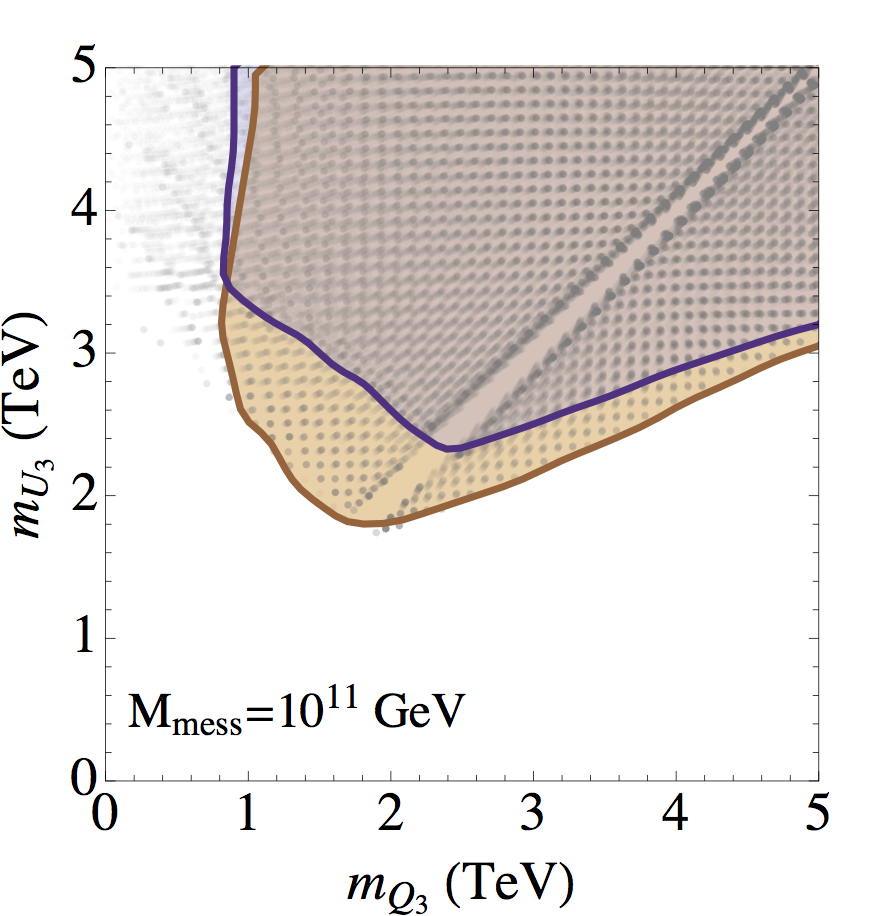}
\\ \vskip0.1cm
\includegraphics[width=.48\textwidth]{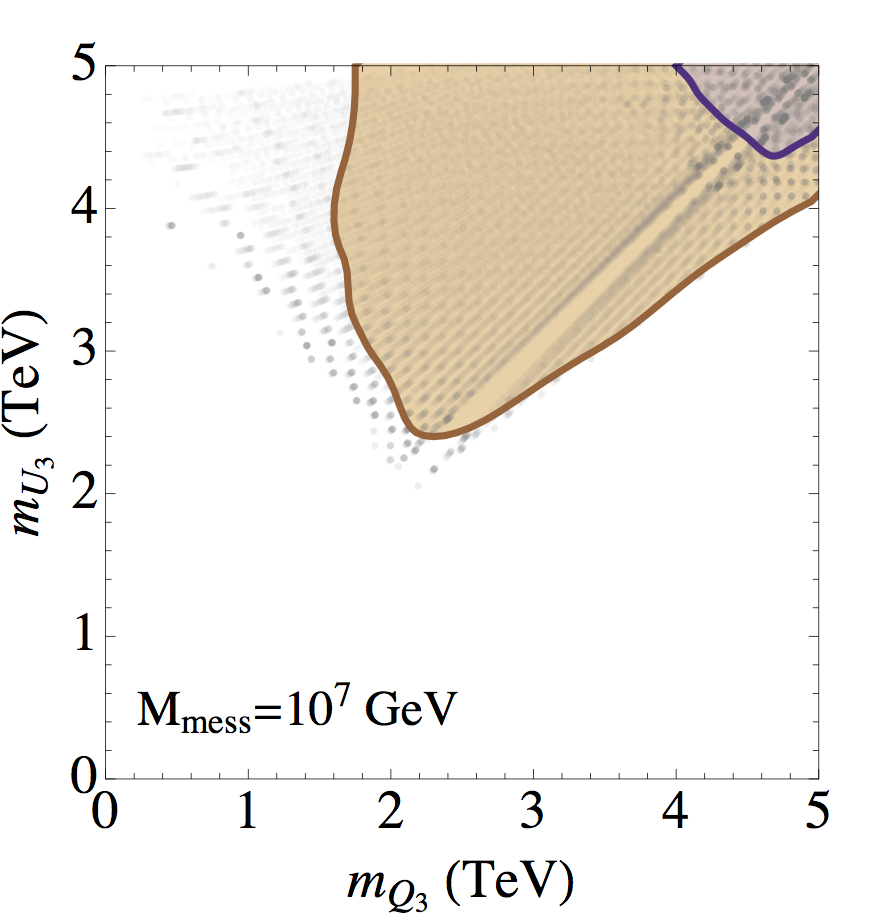}
\caption{Allowed regions in the stop soft mass plane for $\mu<0$ ($\mu>0$) indicated by the blue (orange) shaded regions. The gray dots are the allowed physical stop masses, which can differ significantly from the soft masses due to the gluino threshold correction. The wedge along the diagonal is a result of the level repulsion between the two stop mass eigenstates.}  \label{moneyplane}
\end{figure}

We begin with a ``birds-eye view" of the parameter space: the viable region projected onto the stop mass plane. Shown in fig.~\ref{moneyplane} is the full result of our numerical scan for the three different values of $M_{mess}$ and the two signs of $\mu$. There are several interesting features of these plots which highlight the general points made in section \ref{sec:generalities}. These include:
\begin{itemize}
\item As expected from the discussion in section \ref{subsec:generalpicture}, the allowed region shrinks as $M_{mess}$ decreases.

\item For $M_{mess}=10^{15}$ GeV, the difference between $\mu<0$ (blue) and $\mu>0$ (orange) is minimal, but it becomes increasingly dramatic as $M_{mess}$ decreases. As we will see in more detail below, this is due to the increasing importance of the chargino/neutralino threshold correction to $m_h$.

\item Although the lower bound on $m_{Q_3}$ becomes increasingly stringent with lower $M_{mess}$, the physical mass of the mostly-left-handed stop can be arbitrarily low. We will verify in section \ref{subsec:Qline} that this is due to the gluino threshold correction.

\item The same is not true for the mostly-right-handed stop, whose mass is bounded from below by $\sim (1.5,\, 2,\, 2.5)$~TeV for $M_{mess}=(10^{15},\, 10^{11},\,10^{7})$~GeV. According to (\ref{mEltmU}), the right-handed slepton tachyon constraint prevents $m_{U_3}$ from becoming too light. 

\end{itemize}

\begin{figure}[t]
\centering
\begin{minipage}{0.45\textwidth}
\includegraphics[width=\textwidth]{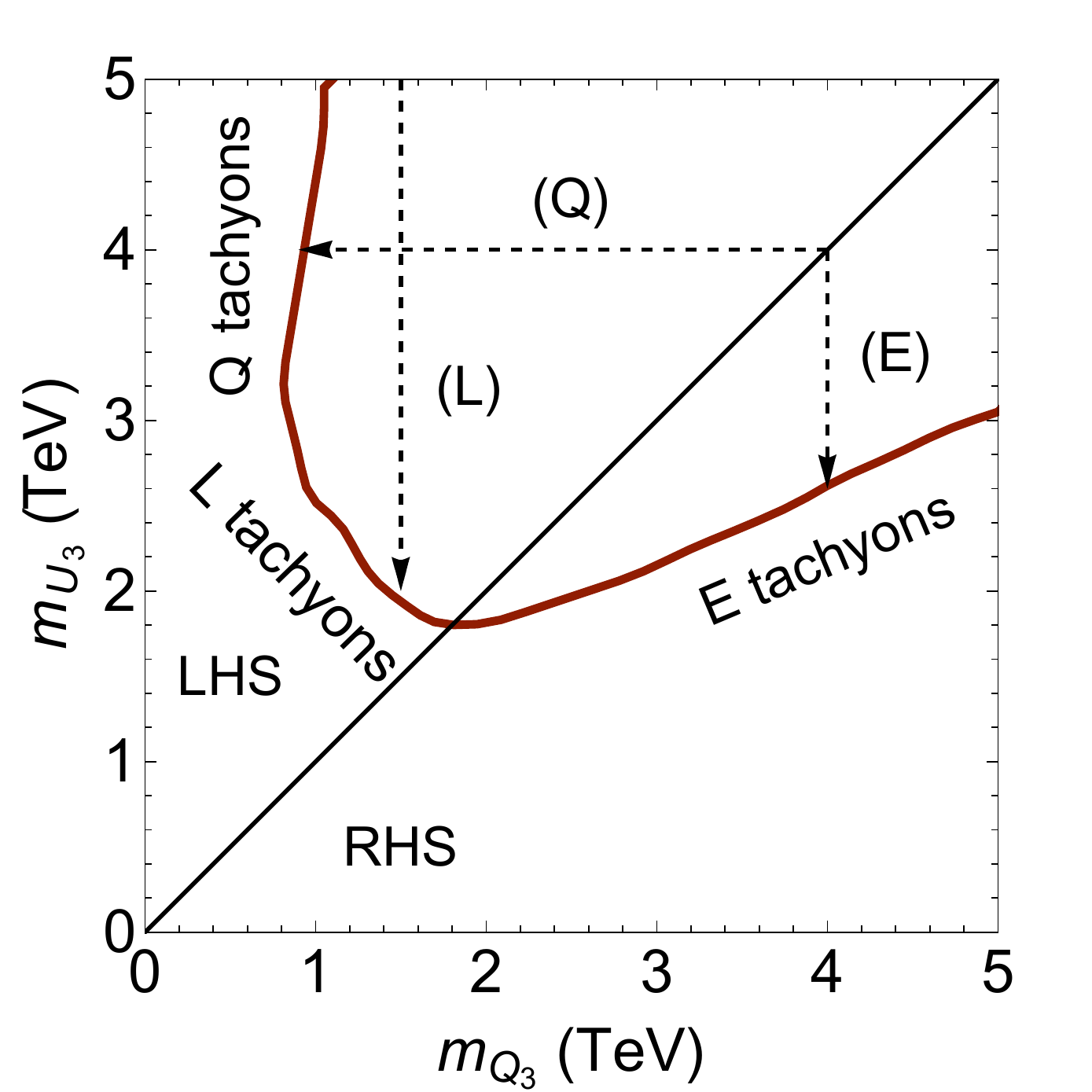}
\end{minipage}
\qquad
\begin{minipage}{0.45\textwidth}
\begin{tabular}{c|c|c}
Line&$M_{mess}$&\\\hline
$L$ &$10^{15}$~GeV &$m_{Q_3} = 1$~TeV\\
$ L $ &$10^{11}$~GeV & $m_{Q_3} = 1.5$~TeV\\\hline
$E$ &$10^{15}$~GeV & $m_{Q_3} = 4$~TeV\\
$E$ &$10^{11}$~GeV & $m_{Q_3} = 4$~TeV\\\hline
$Q$ & $10^{15}$~GeV & $m_{U_3} = 4$~TeV \\
$Q$ & $10^{11}$~GeV & $m_{U_3} = 4$~TeV\\
\end{tabular}
\end{minipage} 
\caption{The same schematic cartoon of the GGM parameter space shown in fig.~\ref{cartoon}, now with three lines labeled by $(Q)$, $(L)$ and $(E)$ indicating the three benchmark lines we study in detail in this section and the next. \label{cartoonlines}}

\end{figure}

In the remainder of this section, we will further elaborate on these and other features by ``zooming in" on these plots and exploring the parameter space along three different benchmark lines depicted in fig.~\ref{cartoonlines}. These lines are chosen in order to illustrate the behavior of the parameter space as we approach the $L$, $E$ and  $Q$ tachyon boundaries. 
Since the allowed parameter space for $M_{mess}=10^7$~GeV is smaller, we will focus on the $M_{mess}=10^{15}$ GeV and $M_{mess}=10^{11}$ GeV cases. 

\subsection{The $L$ and $E$ lines}
\label{subsec:LElines}

These lines are defined by varying $m_{U_3}$ while holding $m_{Q_3}$ fixed to the benchmark values shown in fig.~\ref{cartoonlines}. The physics along the $L$ and $E$ lines is qualitatively similar, so we will consider both simultaneously in this subsection.

\begin{figure}[p]
\centering
\includegraphics[width=0.49\textwidth]{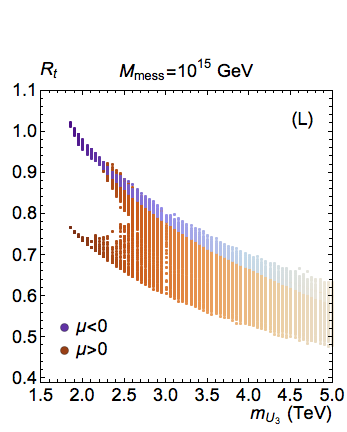}
\hfill
 \includegraphics[width=0.49\textwidth]{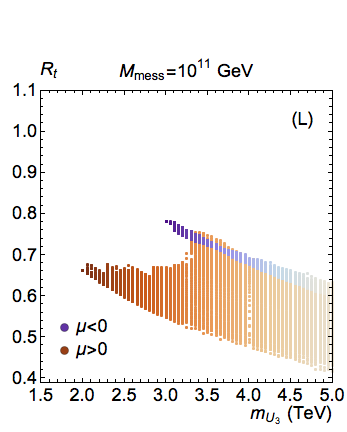}
\includegraphics[width=0.49\textwidth]{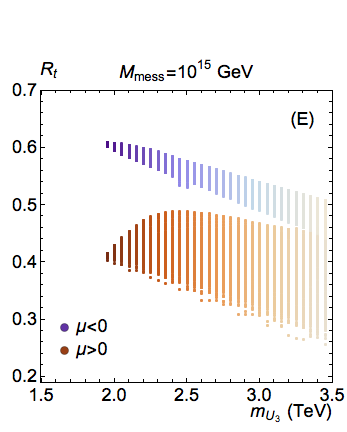}
\hfill
 \includegraphics[width=0.49\textwidth]{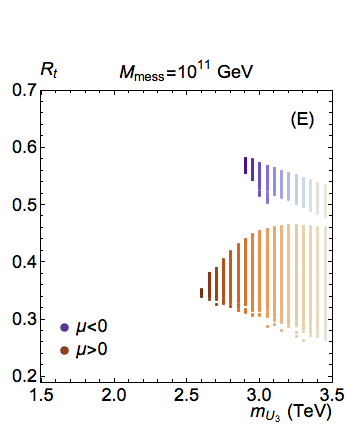}
\caption{$R_t$ as a function of $m_{U_3}$ along the $L$ (top) and $E$ (bottom) benchmark lines, for $M_{mess}=10^{15}$~GeV (left) and $10^{11}$~GeV (right). The orange and blue shaded points correspond to the $\mu>0$ and $\mu<0$ branches respectively. \label{AtvsmU} }
\end{figure}

Shown in fig.~\ref{AtvsmU} are plots of the normalized $A$-term $R_t$ vs. $m_{U_3}$ for these lines in the stop mass plane. As $m_{U_3}$ approaches the boundary (i.e.\ its minimal allowed value), there are two features worth noticing: first,  $R_t$ increases due to the Higgs mass constraint, as expected from fig.~\ref{mhirgrid}. This leads to an increasing tension with EWSB, as explained in section \ref{subsec:generalpicture}. Second, there is a range of $R_t$ values for each $m_{U_3}$, which occurs because we marginalized over the $M_2$ interval in this figure. The range for $R_t$ shrinks to zero once $m_{U_3}$ approaches its minimal allowed value. This indicates that the viable $M_2$ interval shrinks to a point prior to disappearing completely. 

\begin{figure}[t]
\centering
\includegraphics[width=0.49\textwidth]{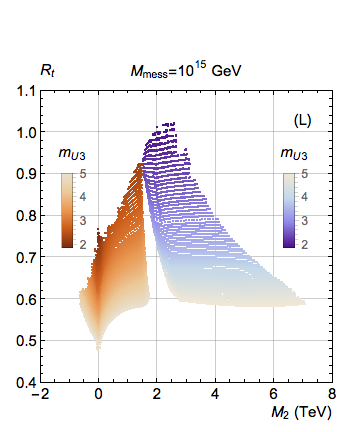}
\hfill
 \includegraphics[width=0.49\textwidth]{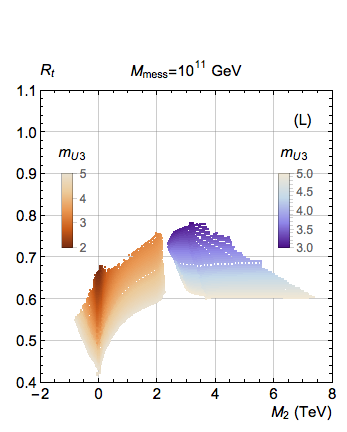}
\includegraphics[width=0.49\textwidth]{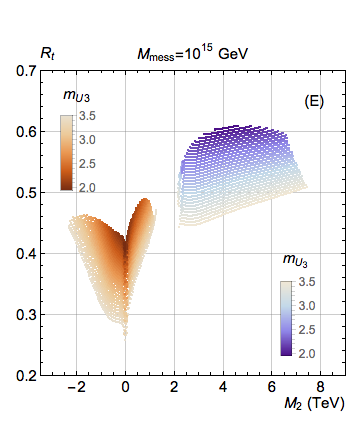}
\hfill
 \includegraphics[width=0.49\textwidth]{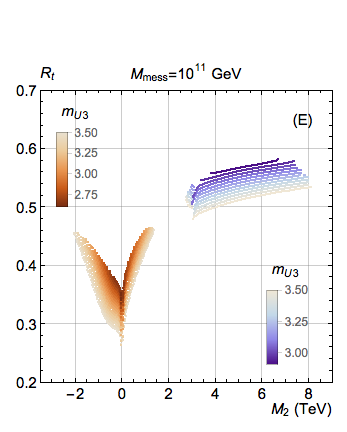}

\caption{Plots of $R_t$ vs.~$M_2$ along the $L$ (top) and $E$ (bottom) benchmark lines, for $M_{mess}=10^{15}$~GeV (left) and $10^{11}$~GeV (right). The orange and blue shaded points correspond to the $\mu>0$ and $\mu<0$ branches respectively. The progressively darker shading of the colors indicates decreasing $m_{U_3}$.
 \label{AtvsM2} }

\end{figure}

Figs.~\ref{AtvsmU} and \ref{AtvsM2} also illustrate very starkly the difference between $\mu<0$ and $\mu>0$: we see that the $A$-terms are mostly constant across the $M_2$ interval for $\mu<0$, as would be expected from the Higgs mass constraint, but for $\mu>0$ they vary quite a lot across the $M_2$ interval.  Evidently, the magnitude of the $A$-term required for $m_h=123$~GeV decreases significantly in the neighborhood of $M_2=0$. 

The reason for this decrease can be traced back to a positive one-loop threshold correction to the Higgs mass coming from loops of light Higgsinos and winos, see appendix  \ref{appChargino/neutralino} for more details.  As the magnitude of $M_2$ decreases, $m_h$ grows by $\sim$~2--3~GeV, and this greatly relaxes the demands on the $A$-term.  Since $M_2=0$ is only accessible for $\mu>0$ due to the pseudoscalar tachyon constraint (first bullet point below (\ref{ewsbir})), only $\mu>0$ is sensitive to this threshold correction. This explains why the allowed parameter space (fig.~\ref{moneyplane}) for $\mu>0$ becomes much larger than the one for $\mu<0$ as $M_{mess}$ decreases. As the constraints on the stop mass plane become more stringent, the importance of the small-$M_2$ threshold correction is magnified. To the point that for $M_{mess}=10^7$~GeV, the constraints basically kill off the entire parameter space, except where this small-$M_2$ threshold correction is present. 

\begin{figure}[t]
\centering
\includegraphics[width=0.49\textwidth]{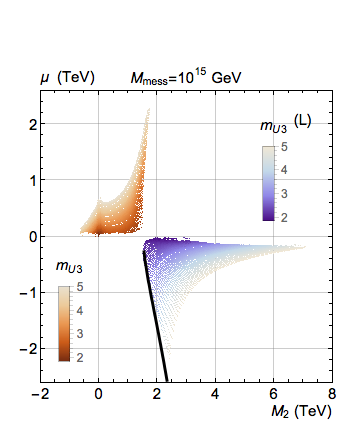}
\hfill
 \includegraphics[width=0.49\textwidth]{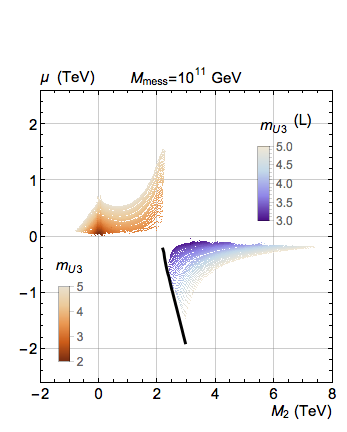}
\includegraphics[width=0.49\textwidth]{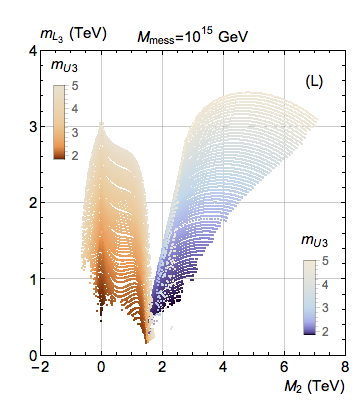}
\hfill
 \includegraphics[width=0.49\textwidth]{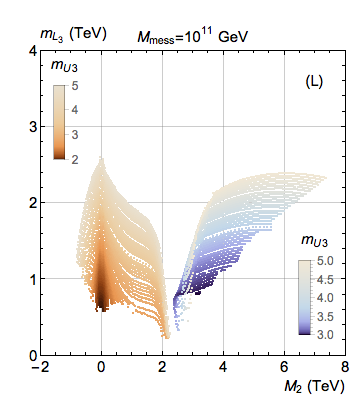}

\caption{Plots of $|\mu|$ (top),  $m_{L_3}$ (bottom)  vs.~$M_2$ along the $L$ benchmark line, for $M_{mess}=10^{15}$~GeV (left) and $10^{11}$~GeV (right). Color schemes are as in fig.~\ref{AtvsM2}. The black curves correspond to the quantitative predictions of the lower end of the $M_2$ interval for $\mu<0$ in the semi-analytic approximation, see   section \ref{sec:interpretation} for details.  }  \label{mumLvsM2}
\end{figure}

\begin{figure}[t]
\centering
\includegraphics[width=0.49\textwidth]{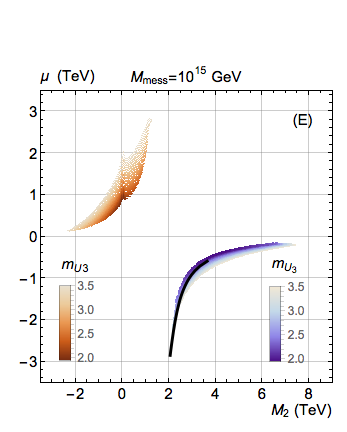}
\hfill
 \includegraphics[width=0.49\textwidth]{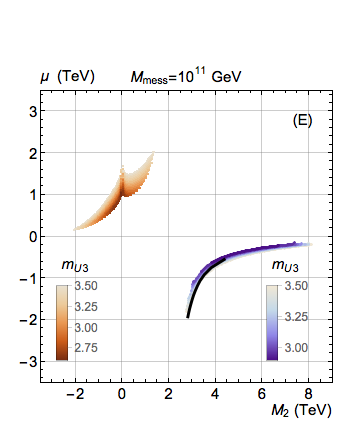}
\includegraphics[width=0.49\textwidth]{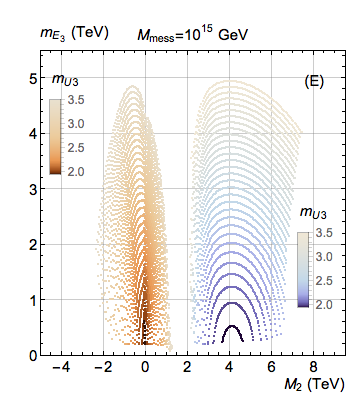}
\hfill
 \includegraphics[width=0.49\textwidth]{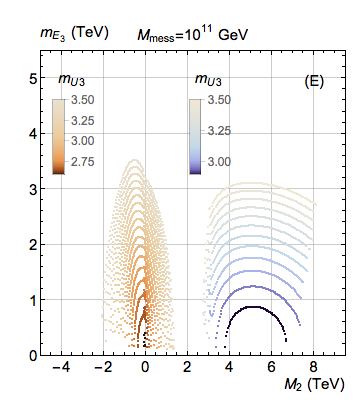}

\caption{Plots of $|\mu|$ (top),  $m_{E_3}$ (bottom) vs.~$M_2$ along the $E$ benchmark line, for $M_{mess}=10^{15}$~GeV (left) and $10^{11}$~GeV (right). Color schemes are as in fig.~\ref{AtvsM2}. The black curves correspond to the quantitative predictions of the lower end of the $M_2$ interval for $\mu<0$ in the semi-analytic approximation, see   section \ref{sec:interpretation} for details. \label{mumEvsM2} }
\end{figure}

Finally, in figs.~\ref{mumLvsM2} and \ref{mumEvsM2} are plots of $\mu$ and the relevant slepton mass across the $M_2$ interval, again with $m_{U_3}$ varying along the $L$ or $E$ line as indicated by the color coding.  In these figures the correlation between the sign of $\mu$ and the viable range of $M_2$ is especially manifest. We highlight some other general features of these plots. For $\mu<0$:
\begin{itemize}

\item $|\mu|$ is always monotonically decreasing with $M_2$.

\item The lower end of the $M_2$ interval is determined by $m_{L_3}\to 0$ or $m_{E_3}\to 0$ on the $L$ or $E$ line respectively, always with $\mu\ne 0$. 

\item The upper end is determined by $\mu\to 0$, and on the $E$ line it is sometimes accompanied by $m_{E_3}\to 0$ (i.e.\ for $M_{mess}=10^{15}$~GeV and lowering $m_{U_3}$ closer to the boundary).  

\end{itemize}
Meanwhile, for $\mu>0$:
\begin{itemize}

\item Again, the $M_2=0$ region has a large effect on the plots. $|\mu|$ is no longer monotonic but tends to rise and fall as we cross $M_2=0$. 

\item In all cases, the allowed $M_2$ interval starts to center around $M_2=0$ as $m_{U_3}$ is lowered. When this happens, the $M_2$ interval becomes bounded by $\mu\to0$ and $m_{E_3}\to 0$ on both ends along $L$ and $E$ lines respectively. 

\end{itemize}
In section \ref{sec:interpretation}, we will understand these features analytically in terms of the approximate IR relations and tree-level EWSB equations described in section \ref{subsec:IRrelations}.

\clearpage

\subsection{The $Q$ line}
\label{subsec:Qline}

\begin{figure}[b!]
\centering
\includegraphics[width=0.49\textwidth]{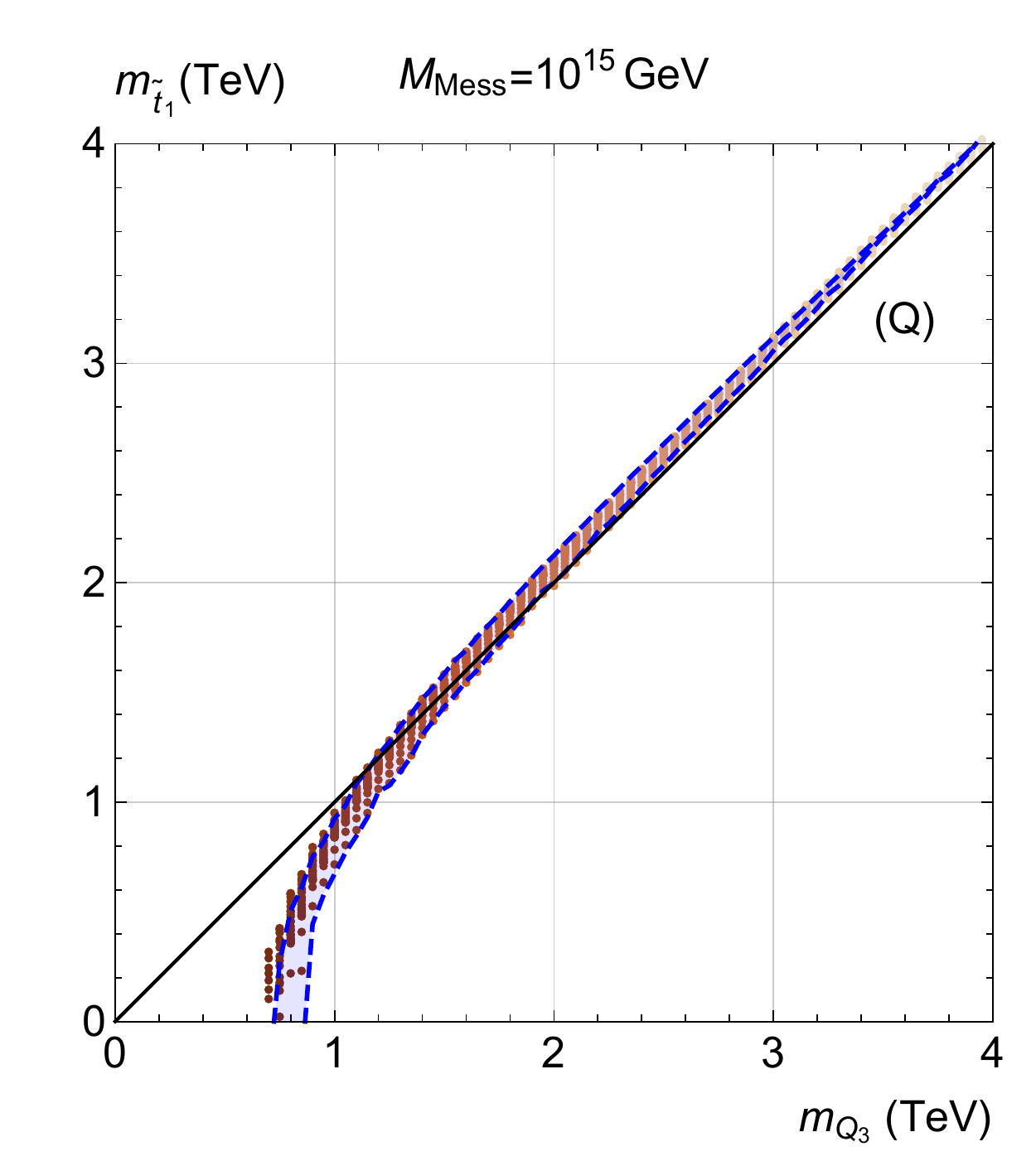}
\hfill
 \includegraphics[width=0.49\textwidth]{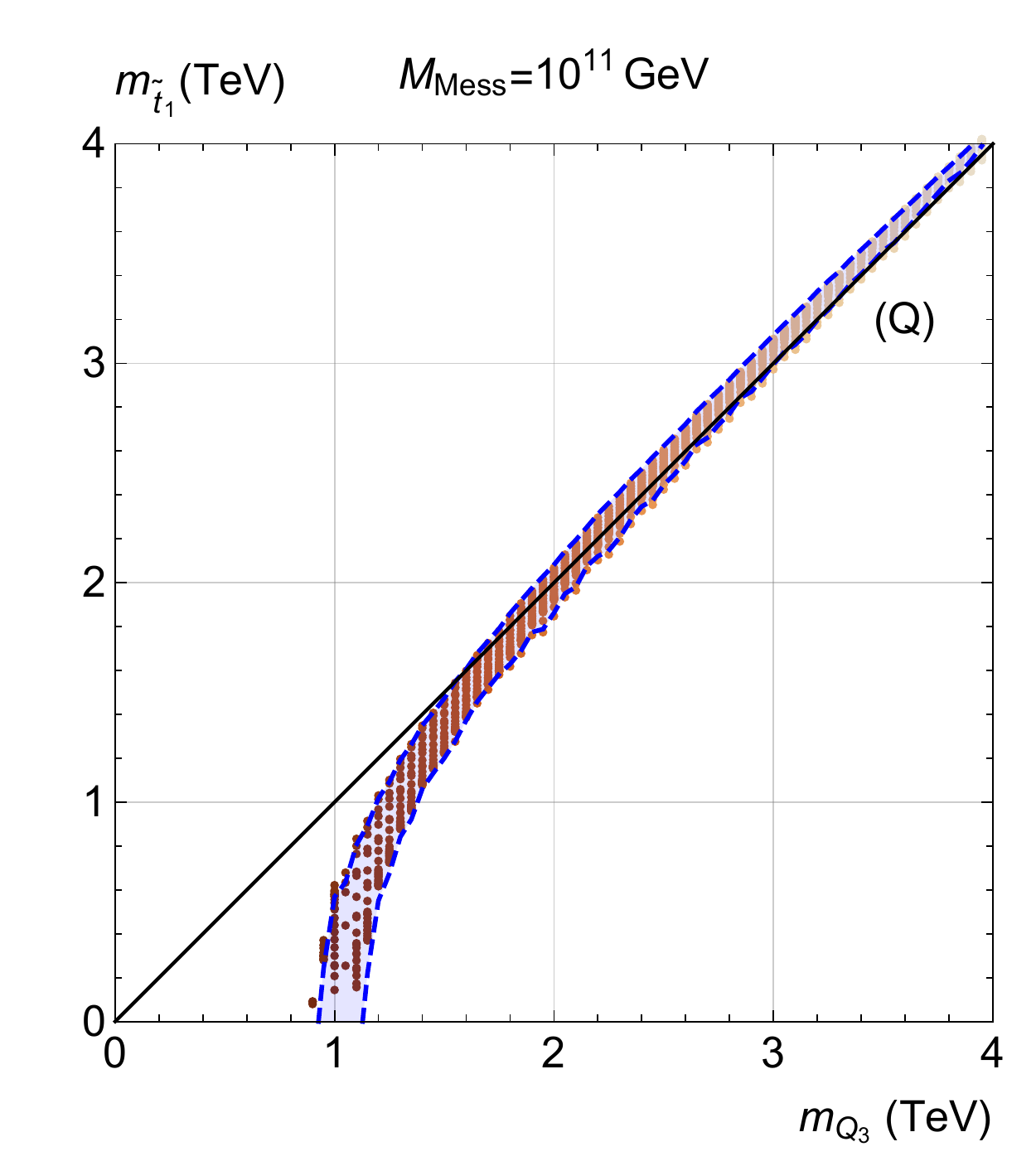}
\caption{Plots of the physical stop pole mass $m_{\tilde{t}_1}$ as a function of $m_{Q_3}$ along the $Q$ benchmark for $\mu>0$. (The $\mu<0$ plots show similar behavior.) The blue shaded region shows the variation within the allowed $M_2$ interval of $m_{Q_3}$ subject to the gluino threshold correction \eqref{bmpzform}.  \label{mst1vsmQ}}
\end{figure}

Finally, we come to the $Q$ benchmark line shown in fig.~\ref{cartoonlines}. Here we fix $m_{U_3}=4$ TeV for both $M_{mess}=10^{15}$ GeV and $M_{mess}=10^{11}$ GeV. The normalized $A$-term, $\mu$ and slepton masses all exhibit the same general behavior along the $M_2$ interval as on the $L$ line, so we will not show these plots again for the $Q$ line. 
The big difference with the $L$ and $E$ lines is that the $Q$ line is not cut off by EWSB and slepton tachyons, but rather by a $Q$ tachyon. Shown in fig.~\ref{mst1vsmQ} is the pole mass of the lightest stop vs $m_{Q_3}$ along the $Q$ line. We see that $m_{\tilde t_1}$  begins to differ significantly from the soft mass $m_{Q_3}$ as we approach the boundary of the stop mass plane, ultimately decreasing to zero. (A similar effect occurs for the pole mass of $m_{\tilde b_1}$.) As in fig.~\ref{moneyplane},  we see that the mostly left-handed stop mass eigenstate can be arbitrarily light despite the Higgs mass constraint.

Also shown in fig.~\ref{mst1vsmQ} is the range across the $M_2$ interval of $m_{Q_3}$ subject to the gluino threshold correction \eqref{bmpzform}. We see that it agrees quite well with the full numerical result given by SoftSUSY. This confirms that the gluino loops dominate the threshold corrections to the lightest stop mass and are ultimately responsible for $m_{\tilde t_1}$ going tachyonic at low $m_{Q_3}$. 

\begin{figure}[t]
\centering
\includegraphics[width=0.49\textwidth]{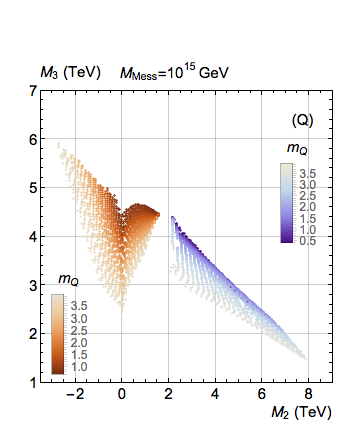}
\hfill
 \includegraphics[width=0.49\textwidth]{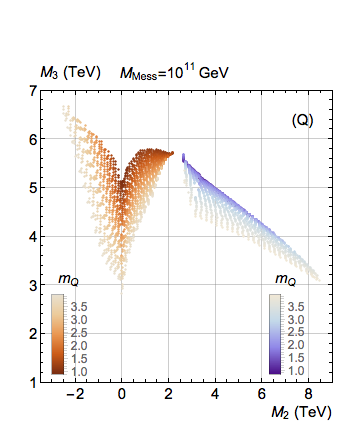}

\caption{Plots of $M_3$ vs.~$M_2$ along the $Q$ benchmark line, for $M_{mess}=10^{15}$~GeV (left) and $10^{11}$~GeV (right). Color schemes are as in fig.~\ref{AtvsM2}.  }  \label{mQline}
\end{figure}

The plots in fig.~\ref{mQline} illustrate more properties of the gluino mass along the $Q$ line.  For $\mu<0$ where the $A$-term is mostly constant, we see the tight linear relationship between $M_2$ and $M_3$ encoded in equation \eqref{gluinosoft}. Furthermore, we see that the Higgs mass constraint forces the gluino mass to be quite large overall, and causes it to grow as $m_{Q_3}$ is lowered. Also the gluino mass is generally larger for $M_{mess}=10^{11}$~GeV than for \mbox{$M_{mess}=10^{15}$~GeV}, since a larger $M_3$ is needed to obtain the desired $A$-term as predicted by equation \eqref{gluinosoft}. The large hierarchy between $m_{Q_3}$ and $M_3$ enhances the gluino threshold correction for low values of $m_{Q_3}$ and lower messenger scales. This is the reason for the increasing lower bound on $m_{Q_3}$ in the plots in fig.~\ref{moneyplane}.

\section{Interpretation}
\label{sec:interpretation}

In this section, we will understand the features of the numerical solution in terms of the tree-level EWSB equations (\ref{ewsbir}) and the IR relations (\ref{IRrelsf}) and (\ref{IRrel1higgs}) . Having achieved an analytical understanding of the $Q$ tachyon boundary through \eqref{bmpzform} in the previous subsection, we will focus on how the EWSB and slepton tachyon boundaries are determined analytically.  

We will organize our discussion in this section around the sign of $\mu$. In previous sections, we have seen repeatedly that the $\mu<0$ and $\mu>0$ branches differ qualitatively due to the presence of the $M_2=0$ threshold correction to the Higgs mass for the latter branch. As a consequence, $A_t$ is basically constant across the $M_2$ interval for $\mu<0$, while this is not the case for the $\mu>0$ branch. For $\mu<0$, this gives us much greater control over the parameter space, since we can fix $A_t$ in all the equations above. The $\mu>0$ branch on the other hand requires greater care, and we will consider it separately. 

For the convenience of the reader, we repeat here the most important formulas and definitions from section \ref{subsec:generalpicture} used in the analysis below. The EWSB equations can be written as 
\bea
& e\,(\delta M_2+d\, A_t)^2 + a\,m_{L_3}^2+\mu^2 \approx m_0^2\\
& - g\,\delta M_2\,\mu\tan\beta \approx m_{L_3}^2+\mu^2
\label{ewsbirrepeat}
\eea
with 
\bea
\delta M_2 &\equiv M_2+f\, A_t\\
m_0^2&\equiv b\,(m_{Q_3}^2+m_{U_3}^2)-c\, A_t^2. \label{deltaM2defrepeat}
\eea
The $\{a,b,c,d,e,f,g\}$ are the numerical constants in table~\ref{tabsumrulesparam} and are determined by the RGEs. The soft mass of the right-handed slepton is furthermore given by
\beq
 m_{E_3}^2 \approx 2m_{L_3}^2+{1\over2}\mu^2 +  {3\over2}(m_{U_3}^2-m_{Q_3}^2).\label{IRrelErepeat}
 \eeq
In the remainder of this section, we will heavily rely on these relations. We further make the following approximations: all of our formulas in this section will be to first non-trivial order in the $1/\tan\beta$ expansion; we are using the tree-level EWSB equations; we are neglecting corrections proportional to $g_1$, $y_b$ and $y_\tau$. Also, for the most part, we will ignore the mild variation of the $\{a,b,c,d,e,f,g\}$ coefficients across the stop mass plane.

\subsection{$\mu<0$: characterizing the $M_2$ interval}

The analysis presented in this subsection and the next applies to points with $\mu<0$, which are the blue shaded points in the plots in section \ref{sec:results}. We begin our discussion in the bulk of the stop mass plane, i.e.\ with large $m_0^2$. Here it is convenient to introduce a new variable:
\beq
m^2\equiv m_0^2 - {3\over4}a(m_{Q_3}^2-m_{U_3}^2)\theta(m_{Q_3}^2-m_{U_3}^2)\label{msqdef}
\eeq
where $\theta$ is the Heaviside step function and $m_0^2$ is defined in \eqref{deltaM2defrepeat}. So $m^2=m_0^2$ on the LHS of the stop mass plane, while it equals $m_0^2-{3\over4}a(m_{Q_3}^2-m_{U_3}^2)$ on the RHS of the stop mass plane.

On the LHS of the stop mass plane, we expect the $M_2$ interval is bounded by left-handed slepton tachyons. Setting $m_{L_3}^2=0$ on the LHS, we find two solutions to (\ref{ewsbirrepeat}), one at small $\delta M_2$:
\bea\label{smallM2lhssol}
& \delta M_2=  {1\over g\tan\beta}\sqrt{ m^2-e\,d^2A_t^2}\\
& \mu =-\sqrt{ m^2-e\,d^2A_t^2}
\eea
and one at large $\delta M_2$:
\bea
& \delta M_2= -d\, A_t + \sqrt{{m^2\over e}}\\
& \mu=0   \label{largeM2lhssol}
\eea
The solutions in (\ref{smallM2lhssol}) and (\ref{largeM2lhssol}) should thus correspond to the two endpoints of the $M_2$ interval.

Sufficiently far into the RHS of the stop mass plane, the $M_2$ interval should be bounded by right-handed slepton tachyons. Setting $m_{E_3}^2=0$ on the RHS again yields two solutions, one at small $\delta M_2$:
\bea
& \delta M_2 = {3\over4}{m^2-e\,d^2 A_t^2+a'(m_{Q_3}^2-m_{U_3}^2)\over g\tan\beta\sqrt{a'(m^2-e\,d^2A_t^2 )}}\\
& \mu = -\sqrt{{m^2-e\,d^2A_t^2 \over a'}}\label{smallM2rhssol}
\eea
and one at large $\delta M_2$:
\bea
& \delta M_2 = -d A_t + \sqrt{{m^2\over e}}\\
& \mu = -{3\over4} {m_{Q_3}^2-m_{U_3}^2\over g\;\delta M_2 \tan\beta}\label{largeM2rhssol}
\eea
with $a'\equiv 1-a/4$. 

In general, the approximate solutions  (\ref{smallM2lhssol}) and (\ref{smallM2rhssol}) correctly characterize the lower endpoint of the $M_2$ interval and the general trends along the $M_2$ interval, but   (\ref{largeM2lhssol}) and (\ref{largeM2rhssol}) fail to characterize the behavior at the upper endpoint of the $M_2$ interval. In more detail:
\begin{itemize}

\item According to the approximate solutions, the lower endpoint of the $M_2$ interval is characterized by $m_{L_3}^2\to 0$ or $m_{E_3}^2\to 0$ with $\mu\ne 0$. These features are all clearly borne out in the full solution, see figs.~\ref{mumLvsM2} and \ref{mumEvsM2}. In these figures, we have also indicated the quantitative predictions of (\ref{smallM2lhssol})  and  (\ref{smallM2rhssol})  for the location of the lower $M_2$ endpoint, as shown by the black line.   We see that it describes the full solution well. 

\item As $\delta M_2$ increases,  it is straightforward to show from the EWSB equations (\ref{ewsbirrepeat}) that $|\mu|$ always monotonically decreases with $\delta M_2$, while $m_{L_3}^2$ and $m_{E_3}^2$ must rise and fall. These trends are clearly borne out in figs.~\ref{mumLvsM2} and \ref{mumEvsM2}.

\item Finally, at the upper endpoint of the $M_2$ interval, the approximate solutions in (\ref{largeM2lhssol}) and (\ref{largeM2rhssol}) predict $m_{L_3}^2\to 0$ or $m_{E_3}^2\to 0$ with $\tan\beta$-suppressed (or zero) $\mu$. While $\mu$ does become quite small in general, we observe that $m_{L_3}^2$ and $m_{E_3}^2$ appear to be cut off at a large value at the upper end of the $M_2$ interval.  This can be traced back to the first EWSB relation in (\ref{ewsbirrepeat}) which implies
\bea
m_{L_3}^2&\approx -g \,\delta M_2\, \mu\tan\beta\\
m_{E_3}^2&\approx -2g\, \delta M_2\,\mu\tan\beta-{3\over2}(m_{Q_3}^2-m_{U_3}^2)
\eea
 when $\mu$ is small.  The factor of $\tan\beta$ and large $\delta M_2$ mean that $\mu$ has to become extremely small before $m_{L_3}$ and  $m_{E_3}$ start to visibly approach zero. Such small values of $\mu$ become sensitive to various effects we have neglected, e.g.\ threshold corrections to the tree-level EWSB equations, and the finite-resolution effects of our grid. Evidently, these are enough to cut out the $m_{L_3}\to 0$ and $m_{E_3}\to 0$ behavior at large $M_2$.

\item  Correspondingly,  we find that the quantitative predictions for the upper endpoint given in  (\ref{largeM2lhssol}) and (\ref{largeM2rhssol})  do not work so well since these assumed $m_{L_3}^2=0$ and $m_{E_3}^2=0$ from the start. 

\end{itemize}

\subsection{$\mu<0$: Approaching the boundaries}

\begin{figure}[t!]\centering
\includegraphics[width=0.49\textwidth]{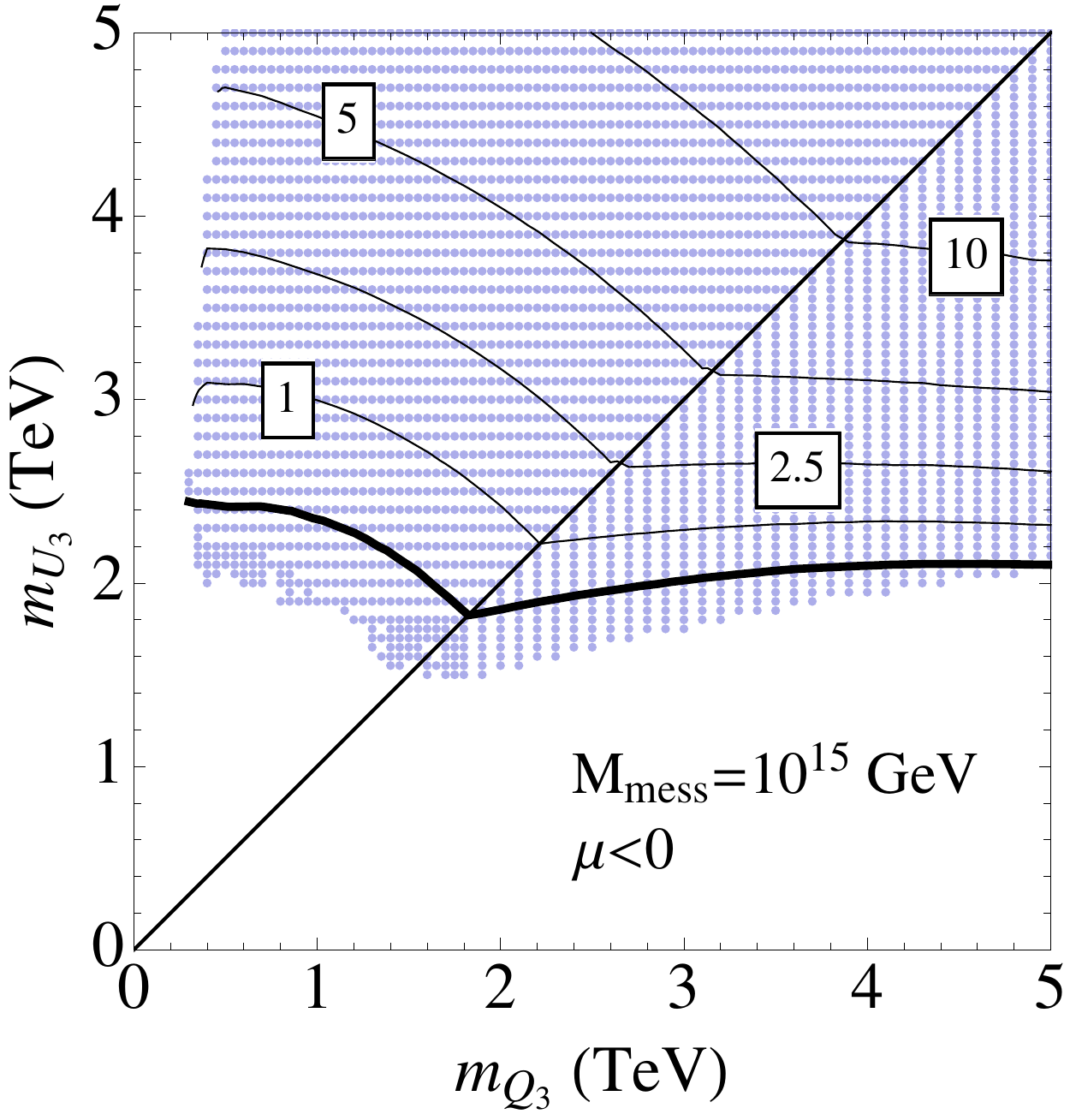}\hfill
\includegraphics[width=0.49\textwidth]{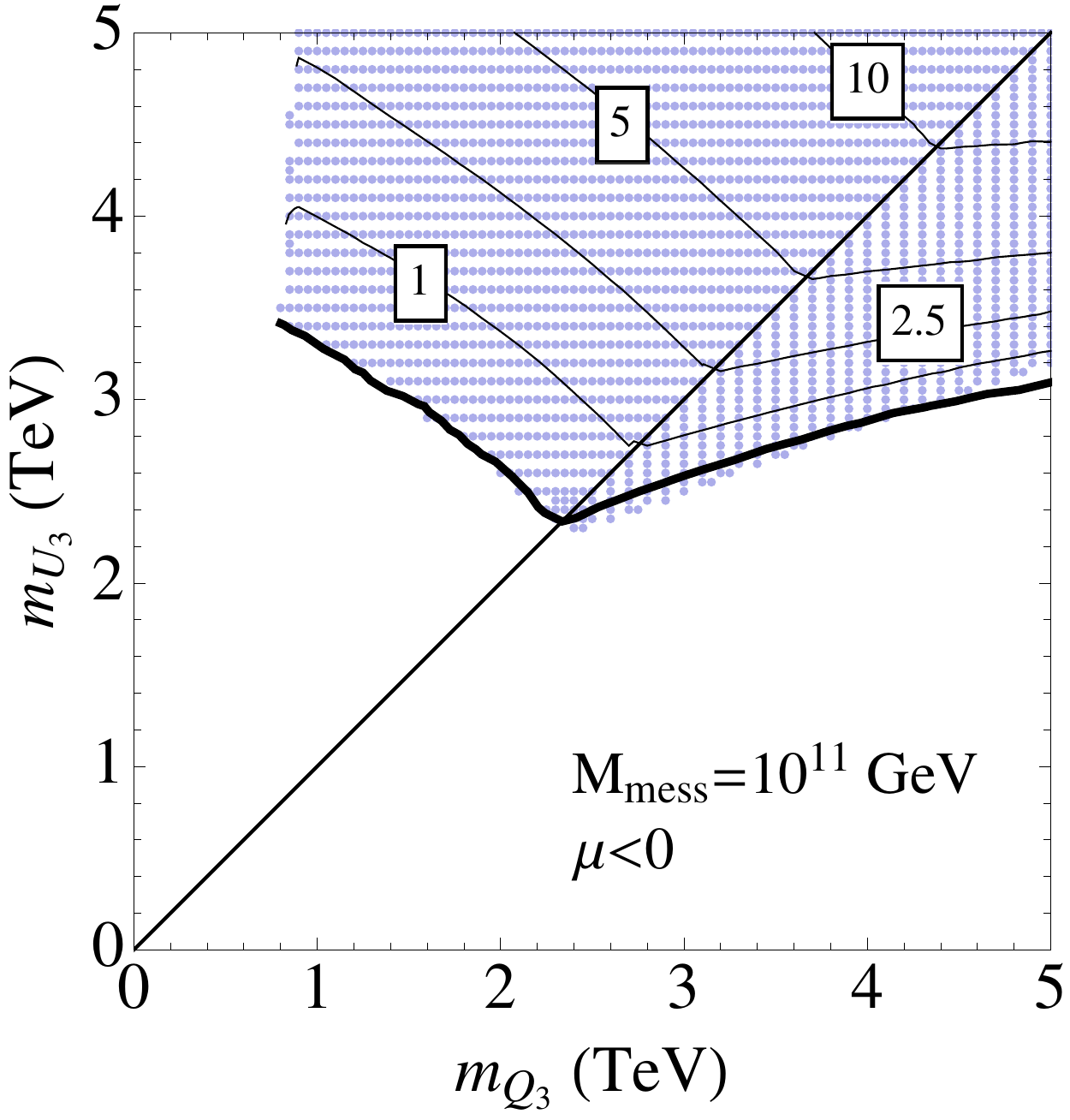}
\caption{Viable points in the stop mass plane for $\mu<0$, with contours of constant $m^2-ed^2 A_t^2$ in $\mathrm{TeV}^2$ (thin) overlaid. The blue dots are the individual points of our full numerical scan, to illustrate our resolution.  We see that the boundary of the stop mass plane is characterized fairly well  by  $m^2= e d^2 A_t^2$ (thick).  }  \label{moneyplaneSAmun}
\end{figure}

Having discussed the behavior of the $M_2$ interval in the bulk of the stop mass plane, now we turn to its behavior as we approach the boundaries of the stop mass plane, i.e.\ as we decrease $m^2$. Shown in fig.~\ref{moneyplaneSAmun} are contours of $m^2-ed^2A_t^2$ for $M_{mess}=10^{15}$ and $10^{11}$~GeV.\footnote{Again, we don't show $M_{mess}=10^7$~GeV here because it is a very tiny region for $\mu<0$ and appears to be subject to threshold corrections from enormous gluino masses that make the semi-analytic approximation unreliable.} At the zero contour, the approximate solutions (\ref{smallM2lhssol}) and (\ref{smallM2rhssol}) become imaginary and are no longer valid. We see that the zero contour does a fairly good job of characterizing the boundary of the stop mass plane.  We have verified that the largest discrepancies for $M_{mess}=10^{15}$~GeV (LHS) arise due to sub-leading effects that we have neglected in this simplified semi-analytic treatment, specifically corrections proportional to $M_1$ and the variation of the transfer matrix along the stop mass plane.
 
We must address one technicality, however, before declaring victory. For $m^2<ed^2A_t^2$, the approximate solutions actually have a second phase where the $M_2$ interval is bounded by $\delta M_2=-dA_t\pm\sqrt{m^2\over e}$. This phase is distinguished by small $\mu$ throughout the $M_2$ interval; in fact, on the LHS, $\mu$ goes to zero at both ends and is non-monotonic on the interval. Because of the very small $\mu$, we expect this entire phase to not be robust against threshold corrections and finite-resolution effects. Indeed, we find that the first phase seems to dominate the parameter space of the full numerical solution, and we only see any evidence for the second phase in a tiny sliver of the LHS of the stop mass plane for $M_{mess}=10^{15}$~GeV. In any event, the question as to whether this phase exists or not is mainly academic, since it would be largely excluded by the LEP bound on charginos.

\subsection{$\mu>0$: the role of the $M_2\approx 0$ region}
\label{subsec:mupboundary}

\begin{figure}[!b]\centering
\includegraphics[width=.45\textwidth]{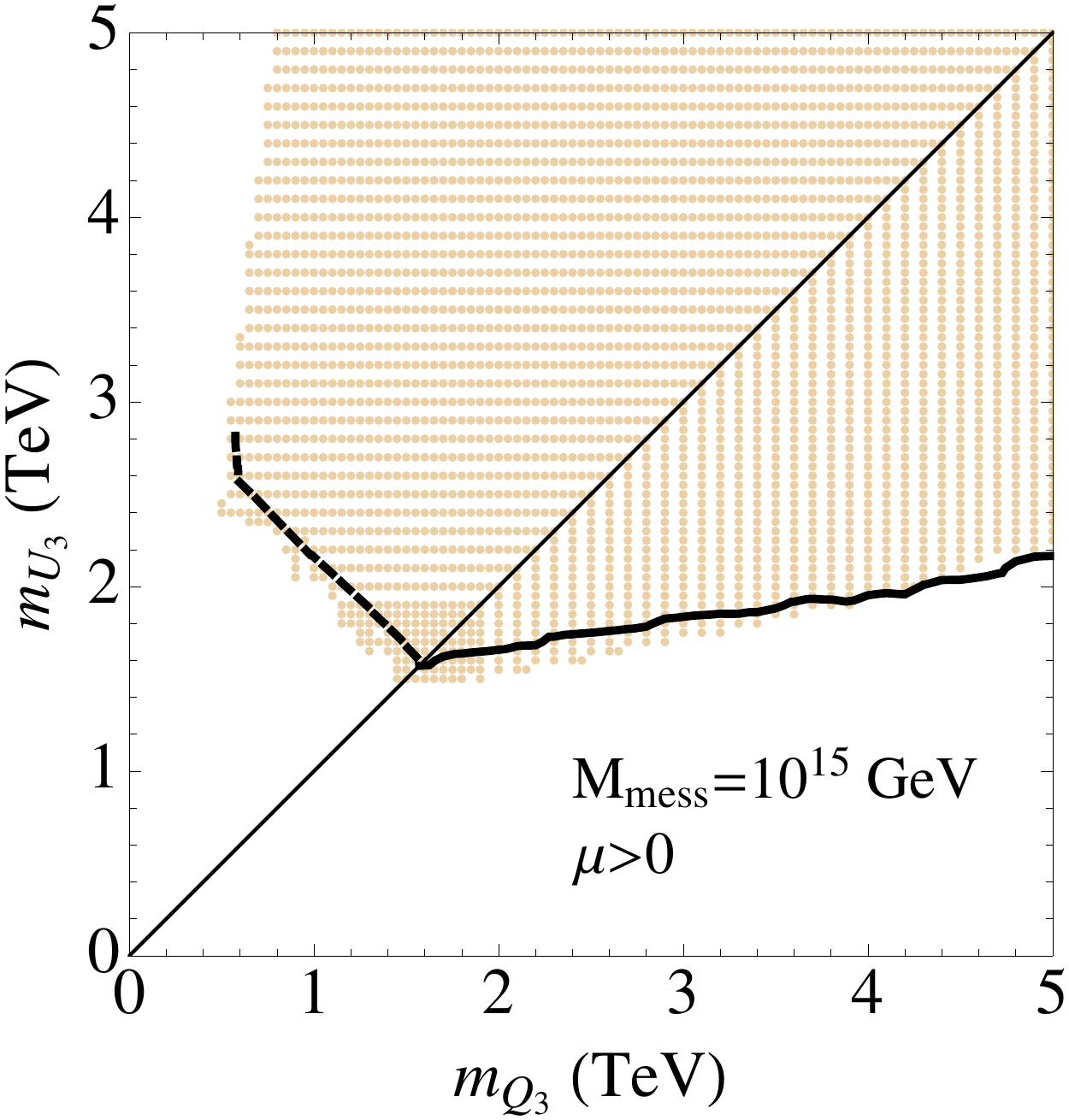}
\quad\includegraphics[width=.45\textwidth]{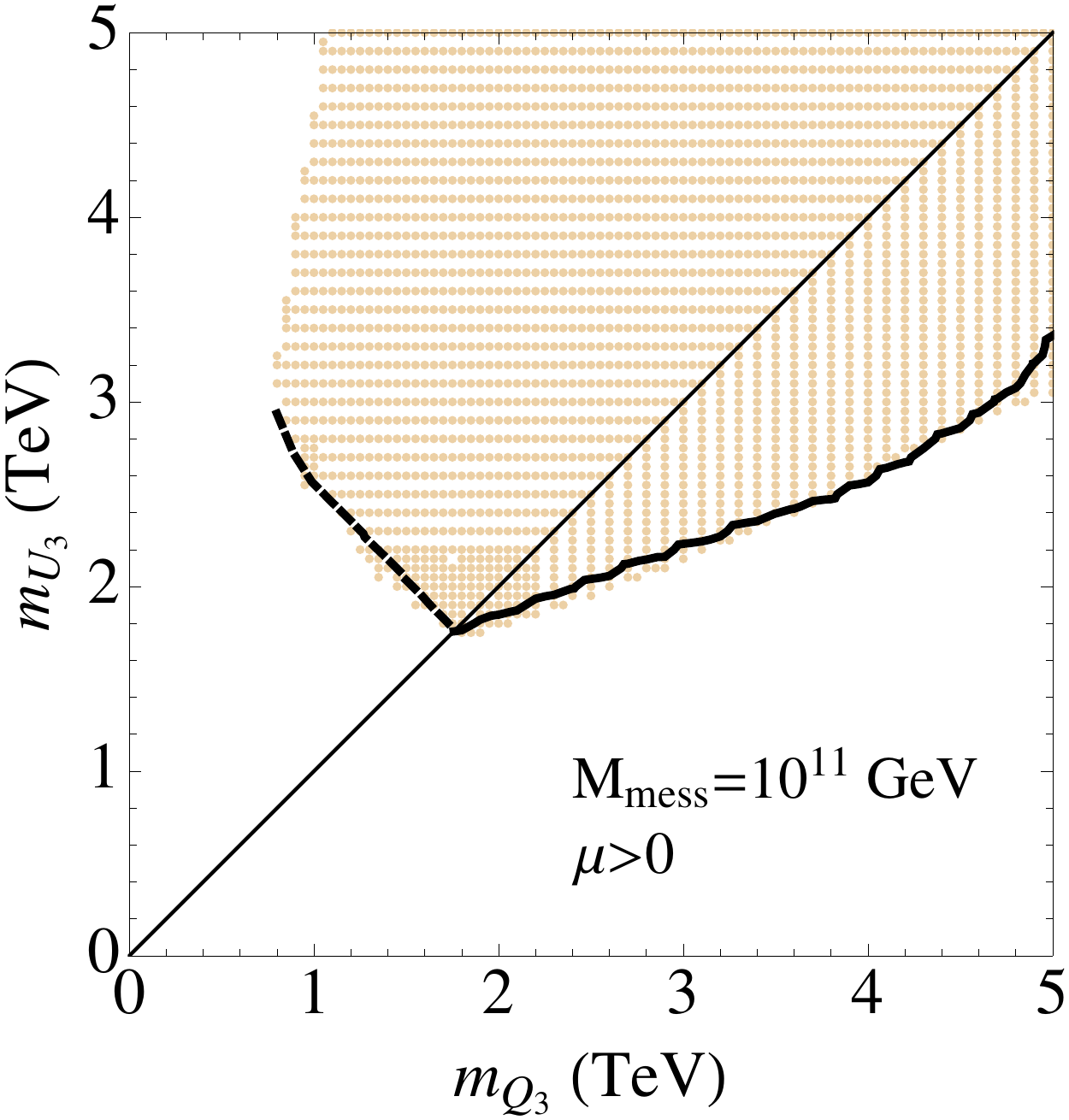}\\
 \vskip0.1cm
\includegraphics[width=.45\textwidth]{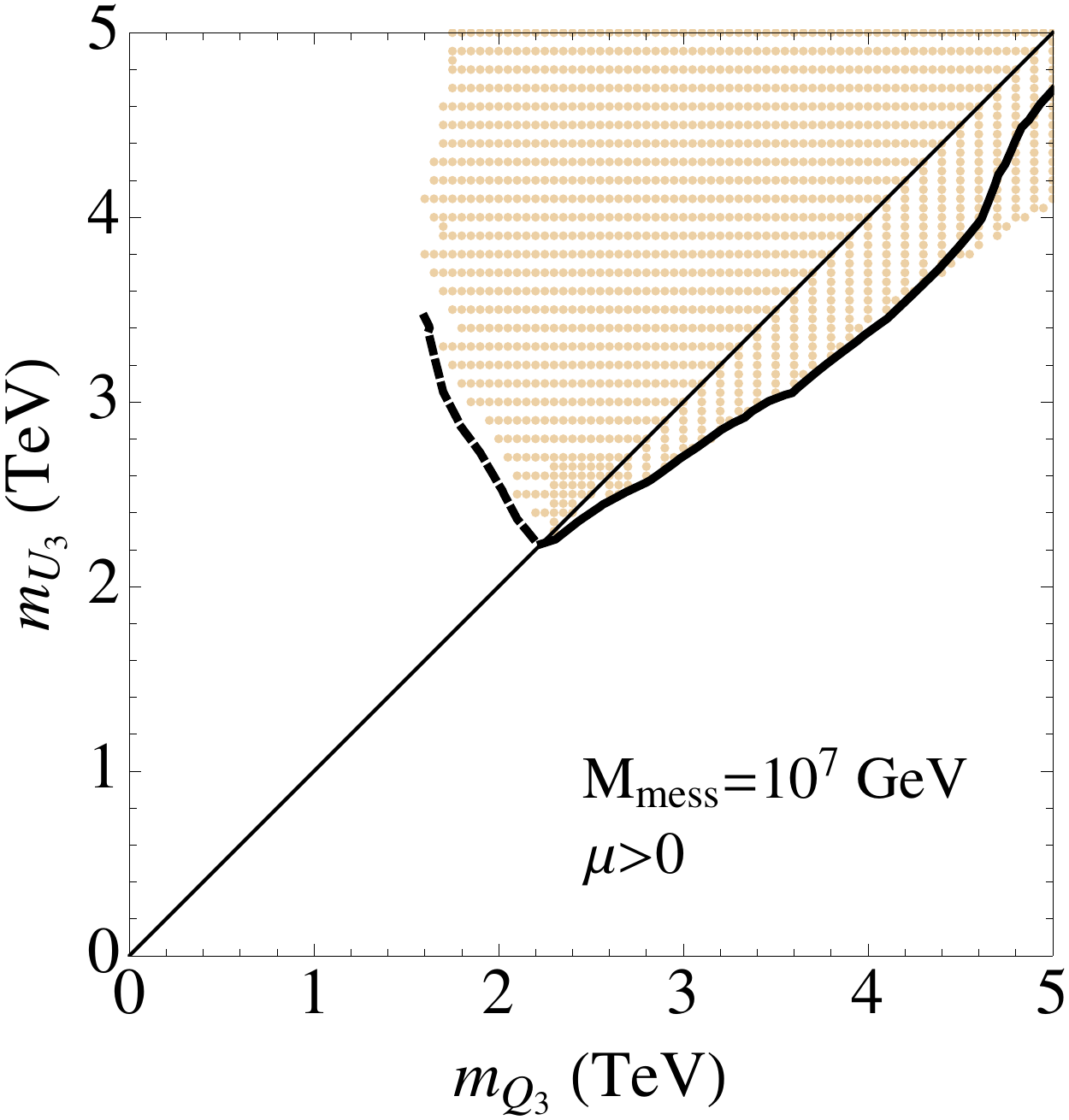}
\caption{Viable points in the stop mass plane for $\mu>0$, with contours of $\mu(M_2=0)=0$ (dashed) and $m_{E_3}^2(M_2=0)=0$ (solid) overlaid. The orange dots are the individual points of our full numerical scan, to illustrate our resolution.}  \label{moneyplanemup}
\end{figure}

Next we turn to the $\mu>0$ case. With just a constant $A$-term, the situation for $\mu>0$ would be nearly identical to that of $\mu<0$. However, we have seen in figs.~\ref{AtvsmU} and \ref{AtvsM2} that the $A$-term needed to achieve~$m_h=123$~GeV depends strongly on $M_2$ in the neighborhood of $M_2=0$. As explained in appendix \ref{appChargino/neutralino}, this is due to the influence of one-loop chargino-neutralino threshold corrections. Since these threshold corrections are positive, the required $A$-term is significantly smaller around $M_2\approx 0$. A smaller $A$-term relieves the tension with tachyons and EWSB, and so this can allow points to survive further into the stop mass plane than would otherwise be the case. We have already seen this illustrated in fig.~\ref{moneyplane}, most dramatically for $M_{mess} = 7$.

Here we will understand this in more detail using the approximate tree-level equations. To begin, let's set $M_2=0$ (i.e.\ $\delta M_2=f A_t$) in (\ref{ewsbirrepeat}). We see that $m_{L_3}^2$ is always large and positive in this regime. Solving for $\mu$ and $m_{E_3}^2$ we find 
\bea
& \mu(M_2=0)= {m_0^2 - e (d+f)^2A_t^2 \over a g f (-A_t)\tan\beta} + \dots\\
& m_{E_3}^2(M_2=0) = 2{m_0^2-e(d+f)^2A_t^2-{3\over 4}a(m_{Q_3}^2-m_{U_3}^2)\over a} + \dots
\label{mumEzerocontour}
\eea
where $\dots$ denote higher order corrections in $1/\tan\beta$. These are monotonically decreasing functions of $-A_t$. As we move away from $M_2=0$, the value of $-A_t$ needed to realize~$m_h=123$~GeV increases significantly, so this has the effect of decreasing $\mu$ and $m_{E_3}^2$. This results in the rise and fall of these parameters around $M_2$  observed in figs.~\ref{mumLvsM2} and \ref{mumEvsM2}. 

As we decrease $m_0^2$, the values of $\mu(M_2=0)$ and $m_{E_3}^2(M_2=0)$ decrease, until eventually they cross zero. Since the $A$-term here is so much smaller than away from $M_2=0$, this can occur further into the stop mass plane than the boundaries discussed in the previous subsection. In fig.~\ref{moneyplanemup} we show the allowed points for $\mu>0$, with the zero contours of $\mu(M_2=0)$ and $m_{E_3}^2(M_2=0)$ overlaid as predicted by the semi-analytic method. (To improve the accuracy of these contours, we have included the sub-leading $1/\tan\beta$ corrections in (\ref{mumEzerocontour}).) We see that this does an excellent job of characterizing the boundary of the stop mass plane for $\mu>0$.

\section{Discussion\label{sec:conclusions}}
\subsection{Summary}
\label{subsec:summary}

In this paper, we have initiated a comprehensive study of the GGM parameter space following the discovery of the Higgs at $m_h=125$~GeV. In pure GGM, we have at the messenger scale: flavor universality, three independent gaugino masses, $B_\mu=A_t=0$, and $\mu$ ``set by hand". Using a transfer matrix approach to the MSSM RGEs, we traded the GGM parameter space defined at the messenger scale $M_{mess}$ for a set of IR variables $(m_{Q_3},\,\, m_{U_3},\,\, m_{L_3},\,\, M_1,\,\, M_2,\,\, A_t,\,\, \mu)$.  The RGEs depend only weakly on $M_1$ through the hypercharge coupling, and the results do not qualitatively depend on it. We therefore fixed $M_1=1$~TeV throughout the analysis. We furthermore chose three benchmark values of $M_{mess}=10^{15}$, $10^{11}$, $10^{7}$~GeV. Then the EWSB and Higgs mass constraints eliminated three variables, leaving us to scan over $(m_{Q_3},\,\,m_{U_3},\,\,M_2)$. At each point in the stop mass plane, the viable parameter space was an interval in $M_2$. 

We performed a detailed numerical scan  and validated it analytically using a set of approximate IR relations together with the tree-level EWSB conditions. Through these methods, we thoroughly explored the physically viable parameter space, with a special focus on how its boundaries are determined. By using the IR relations to express all other IR soft masses in terms of (\ref{parameterspace}), we showed that the only relevant constraints on the GGM parameter space (besides EWSB and the Higgs mass) were slepton tachyons and left-handed stop/sbottom tachyons. All other scalar tachyons were less constraining. Furthermore, we showed that for $m_{Q_3}<m_{U_3}$, only left-handed slepton tachyons and left-handed stop/sbottom tachyons were relevant, while for $m_{Q_3}>m_{U_3}$ only right-handed slepton tachyons were relevant. Finally, we saw that the constraints became more stringent and the viable parameter space smaller as $M_{mess}$ was lowered.

One of the most striking results of our analysis was an absolute lower bound on right-handed stop mass, coming from a combination of EWSB and slepton tachyon constraints. In particular $m_{U_3}\gtrsim\,1.5,\, 2,\, 2.5$~TeV for  $M_{mess}=10^{15}$, $10^{11}$, $10^{7}$~GeV respectively. However, for any messenger scale, the left-handed stop and sbottom could be arbitrarily light. The constraints were always such that the optimal point for $m_h=125$~GeV with \mbox{$m_{Q_3} \sim m_{U_3}\sim |A_t|/\sqrt{6}\sim 1$ TeV} could not be achieved in GGM.

We also identified the sign of $\mu$ as playing an important role in the qualitative behavior of the parameter space. For $\mu<0$, the $A$-term was mostly constant across the $M_2$ interval, being determined by the Higgs mass constraint. However, for $\mu>0$, a positive threshold correction to $m_h$ coming from light charginos and neutralinos allowed the $A$-term to be much smaller in the neighborhood of $M_2\approx 0$. This played an especially important role for $M_{mess}=10^7$~GeV, where the tension between the Higgs mass constraint and EWSB and tachyons was so strong that essentially the only viable parameter space had $\mu>0$ and $M_2\approx 0$.

\subsection{Preview of the LHC phenomenology}
\label{subsec:preview}

In a companion paper \cite{GGMcollider}, we will explore the LHC phenomenology of the GGM parameter space with $m_h=125$~GeV. In this subsection we will give a brief preview.

Our semi-analytic understanding of the GGM parameter space gives us great control over the spectrum as we move around in the stop mass plane and the $M_2$ interval.   In particular, it allows us to understand under which conditions a given SUSY particle can be light. On the one hand, this gives us sharp predictions for the properties of the NLSP, and on the other hand it singles out the dominant production channels. Together these two pieces of information  determine most of the collider phenomenology. 

We saw in our analysis how the constraints became increasingly more stringent as $M_{mess}$ was decreased. Assuming the conventional relation between NLSP lifetime and the messenger scale (see e.g.\ \cite{Giudice:1998bp} for a review), this implies a strong preference for long-lived NLSPs at the LHC. While long-lived neutral NLSPs escape the detector without leaving any track, long-lived NLSPs carrying SM charges are very well constrained at the LHC by inclusive CHAMP searches. These bounds will play a substantial role in constraining the GGM parameter space. For the lowest value of $M_{mess}$ that we considered  ($M_{mess}=10^{7} \text{ GeV}$), the NLSP decay to the gravitino may be non-prompt but  still inside the detector volume. Constraining these scenarios is an interesting challenge for LHC searches (see \cite{Liu:2015bma} for a recent discussion) and our work further motivates efforts to improve coverage at Run~II.

The dominant component of the colored production cross section will come from left-handed squarks throughout much of the GGM parameter space. We showed already that the left-handed stop/sbottom can be arbitrarily light. The IR relations (\ref{IRrelsf}) indicate that the left-handed squarks of the first and second generations are heavier, but there are points on the $M_2$ interval where $\mu\sim m_{L_3}\sim 0$, so they can also become very light. This can be verified in our full numerical scan. Meanwhile, the gluinos are generally forced to be very heavy by the Higgs mass constraint, especially at lower stop masses and/or lower messenger scales. (Of course, with sufficiently heavy stops, $A$-terms are not required for the Higgs mass, and then the gluino can be arbitrarily light.)  Finally we saw how the right-handed stops are always forced to be at least 1.5~TeV due to the right-handed slepton tachyon constraint. The IR relations (\ref{IRrelsf}) imply that the 1st/2nd generation right-handed up squarks are even heavier. A more detailed study of the parameter space reveals that the same is true for the right-handed down squarks.

Light EW superpartners are a generic feature of the GGM parameter space. For example, we have seen that light Higgsinos in conjunction with light left or right-handed sleptons always accompany the $L$ and $E$ boundaries of fig.~\ref{cartoon} respectively. If $\mu>0$ we also expect a light wino throughout much of the parameter space, since the Higgs mass constraint selects out the neighborhood of  $M_2\approx 0$. This feature is especially important for low messenger scales, where the $\mu<0$ branch does not allow for light stops. Finally, due to the IR relation 
\begin{equation}\label{IRrelA}
m_A^2\approx m_{L_3}^2+\mu^2, 
\end{equation} 
the pseudoscalar may also be light. (Note that (\ref{IRrelA}) holds anywhere in the parameter space, and will be strong test of GGM should these particles all be discovered.) The prevalence of all these light EW sparticles in GGM, often accompanied by decoupled colored sparticles, provides further motivation for dedicated  Run II searches of direct EW superpartner production.  

Since the boundaries of the parameter space were determined by the combination of the Higgs mass, EWSB, and a tachyon, the GGM spectrum becomes especially predictive here. The tight connection between light EW states and the lightest possible $m_{U_3}$ for a given $M_{mess}$ has a number of important consequences for collider searches. In particular, LEP bounds on EW states indirectly provide a lower bound on $m_{U_3}$.  A future lepton collider such as ILC is expected to further probe a very large portion of the low $m_{U_3}$ region of the GGM parameter space.

\subsection{Future directions}
\label{subsec:futuredirections}

We conclude by discussing some future directions. First, an important question   is to what extent  the constraints we derived here depend on the particular structure of the GGM boundary conditions at $M_{mess}$.  Here we briefly comment on more general scenarios: 

\begin{itemize}

\item The minimal extension of GGM relaxes the requirement of messenger parity, allowing for $U(1)_Y$ $D$-tadpoles. This possibility was already discussed in \cite{Meade:2008wd}; see \cite{Dimopoulos:1996ig,Argurio:2012qt} for explicit weakly coupled realizations. This breaks one of the sum rules in \eqref{sumrules}, leaving the residual ones:
\bea
&m_{H_d}^2= m_{L}^2\\
&m_Q^2+3 m_U^2-9m_D^2-6m_L^2+m_E^2=0\\
&2m_Q^2-3 m_U^2+3 m_L^2-2m_E^2-6 m_{H_u}^2=0\label{sumrulesDtad}
\eea
Consequently, an additional parameter must be added to the list \eqref{UVparameterspace}.
Full control over the resulting  $8$ dimensional parameter space might still be feasible by applying a similar strategy to the one we used here. The result can be interesting since the direct relation between $m_{H_u}$ and $m_L$ induced by \eqref{sumrules} is now broken by the D-tadpole contributions. As a consequence the friction between large $A_t$ and EWSB which was at the basis of our reasoning might be considerably alleviated and lighter stop masses could be viable.

\item  Adding a flavor blind mechanism to generate $\mu$ and $B_\mu$ will generically break the sum-rules \eqref{sumrules} in a model dependent way which cannot be parametrized by a reduced set of sum-rules like \eqref{sumrulesDtad}. Non-zero $A_t$ may also be generated at $M_{mess}$, which would obviously fundamentally alter the nature of the Higgs mass constraint.
None of our conclusions can be then directly extrapolated to extended gauge mediation scenarios such as those in \cite{Evans:2011bea,Evans:2012hg,Kang:2012ra,Craig:2012xp,Abdullah:2012tq,Kim:2012vz,Byakti:2013ti,Craig:2013wga,Evans:2013kxa,Calibbi:2013mka,Jelinski:2013kta,Galon:2013jba,Fischler:2013tva,Knapen:2013zla,Ding:2013pya,Calibbi:2014yha,Basirnia:2015vga}. It may however be possible to perform a similar model independent analysis by making use of the framework developed in \cite{Komargodski:2008ax,Craig:2013wga}. 

\item One can still focus on extended gauge mediation scenarios where $A_t$ is suppressed at $M_{mess}$ and is purely generated by RG evolution. This happens for example in solutions of the $\mu$/$B_\mu$ problem which involve Higgs interactions with heavy singlets \cite{Komargodski:2008ax}. $A_t$ can also be suppressed by an appropriate discrete $R$-symmetry \cite{Backovic:2015rwa}. In this context it would be interesting to account for the extra UV contributions to $m_{H_u}^2$ and $m_{H_d}^2$ along the lines of what we have done here. As in the hypercharge $D$-term scenario, such contributions could alleviate the tension between EWSB and light stops, and possibly allow for the optimal point of $m_h=125$~GeV with $m_{Q_3} \sim m_{U_3}\sim |A_t|/\sqrt{6}\sim 1$ TeV.

\end{itemize}

Another interesting direction for the future would be to study other aspects along the GGM parameter space which are not directly related to collider searches:

\begin{itemize}
\item One of the peculiar features of GGM is that large $A_t$ can only be achieved via large $M_3$ and therefore light stops require a careful tuning of the UV soft masses against the gluino RGE contribution. We expect this extra source of fine-tuning to play a significant role in the tuning measure. While we did not attempt to do so in this paper it would be interesting to quantify the tuning in GGM and comparing it against other UV complete gauge mediation models like the ones in \cite{Evans:2013kxa}. 

\item In order to compensate for the effect of heavy gluinos, the stops run tachyonic shortly above the weak scale \cite{Draper:2011aa}. Moreover, depending on the region of the parameter space, other UV tachyons are necessary in order to obtain EWSB in the IR. Thus the electroweak vacuum tends to be metastable. The  estimates  in \cite{Riotto:1995am} suggest that the vacuum decay is not a stringent constraint, but it would be interesting to perform a careful analysis in these GGM scenarios.

\item Cosmological bounds can also play an important role in GGM parameter space. In order to avoid gravitino overabundance and possible dangerous effects of the NLSP decays on the BBN products, inflation at particularly low temperature is required (see for example \cite{Giudice:1998bp} and references therein for a discussion of the GGM cosmology). This bound on the reheating temperature can be evaded for example by adding tiny RPV couplings. However, it would be interesting to take it seriously and investigate in full generality the allowed cosmological scenarios in GGM. 
\end{itemize}

Finally, let us mention that our procedure has some intrinsic limitations  due to the theoretical uncertainty in the Higgs mass computation. In this paper, we imposed $m_h=123$~GeV in order to optimistically account for this uncertainty. It will be important to revisit this work after future improvements to the accuracy of the Higgs mass calculation, especially if these turn out to contribute negatively to $m_h$.  (See for example the recent discussion in \cite{Vega:2015fna}.) Aside from the usual corrections from higher orders and uncertainties in SM inputs like $m_t$, the Higgs mass computation in GGM is particularly challenging due to the large hierarchies that are present in the colored spectrum.  Perhaps the most acute example of such a situation is given by the $Q$ boundary of our parameter space, where $m_U\gg m_Q$ and also the gluino mass is very large. Using effective field theory techniques such as in \cite{Espinosa:2001mm}, it would be interesting to have a better control on the Higgs mass computation in such a scenario where the lightest stop masses are realized in GGM.

\section*{Acknowledgements:}
We thank Nathaniel Craig, Patrick Draper,  Daniel Egana-Ugrinovic, Alberto Mariotti, David Morrissey, Michele Papucci, Nausheen Shah, Pietro Slavich, Giovanni Villadoro, Hojin Yoo and Kathryn Zurek for interesting discussions. We are also grateful to Alberto Mariotti and Robert Ziegler for comments on the manuscript. The work of SK was supported by the LDRD Program of LBNL under U.S. Department of Energy Contract No. DE-AC02-05CH11231. The work
of DR was supported by the ERC Higgs at LHC. The work of DS was supported in part by DOE grant No.\ DE-SC0003883.

\appendix

\section{Validation plots}\label{appALGO}

In this section we will exhibit some plots validating the accuracy of the transfer matrix and the numerical algorithm for solving $B_\mu({\rm UV})=0$ and $m_h=123$~GeV.

\begin{figure}[h!]

\RawFloats

\begin{minipage}{\textwidth}
\subfigure{\includegraphics[width=0.45\textwidth]{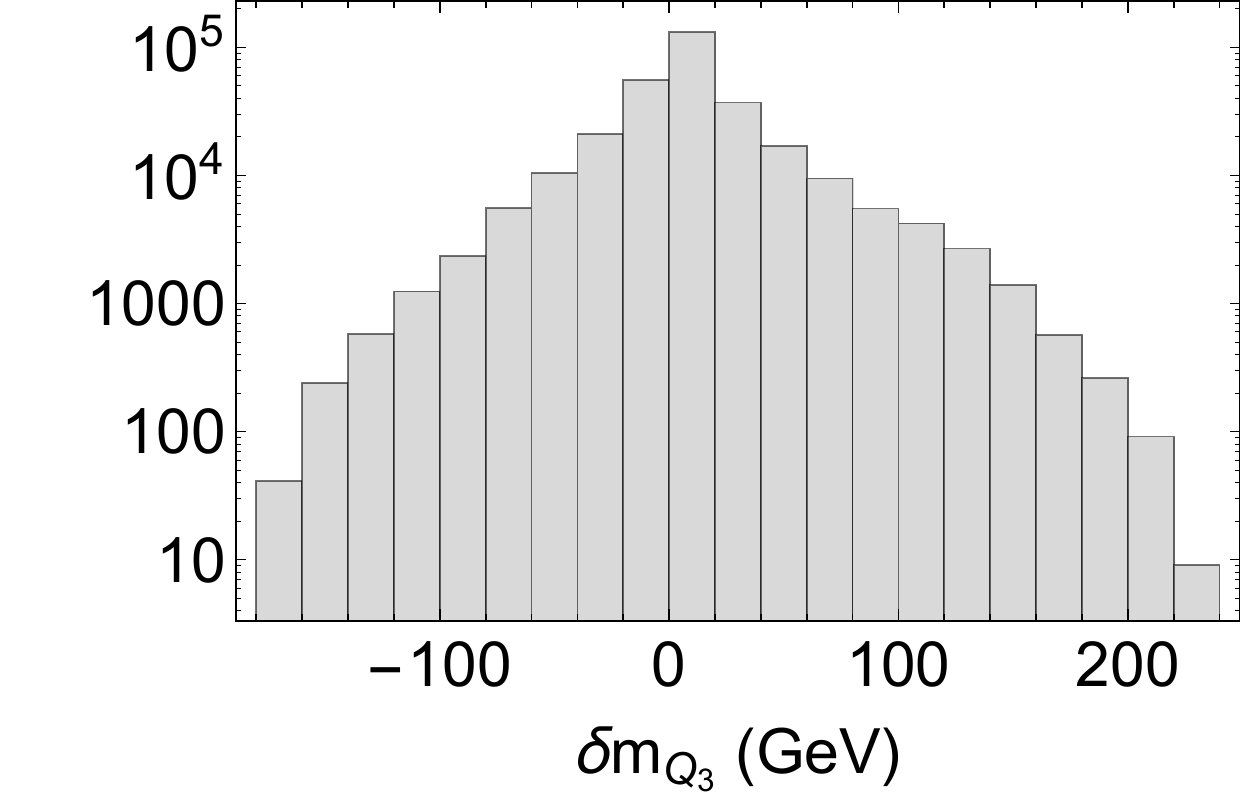}}\hfill
\subfigure{\includegraphics[width=0.45\textwidth]{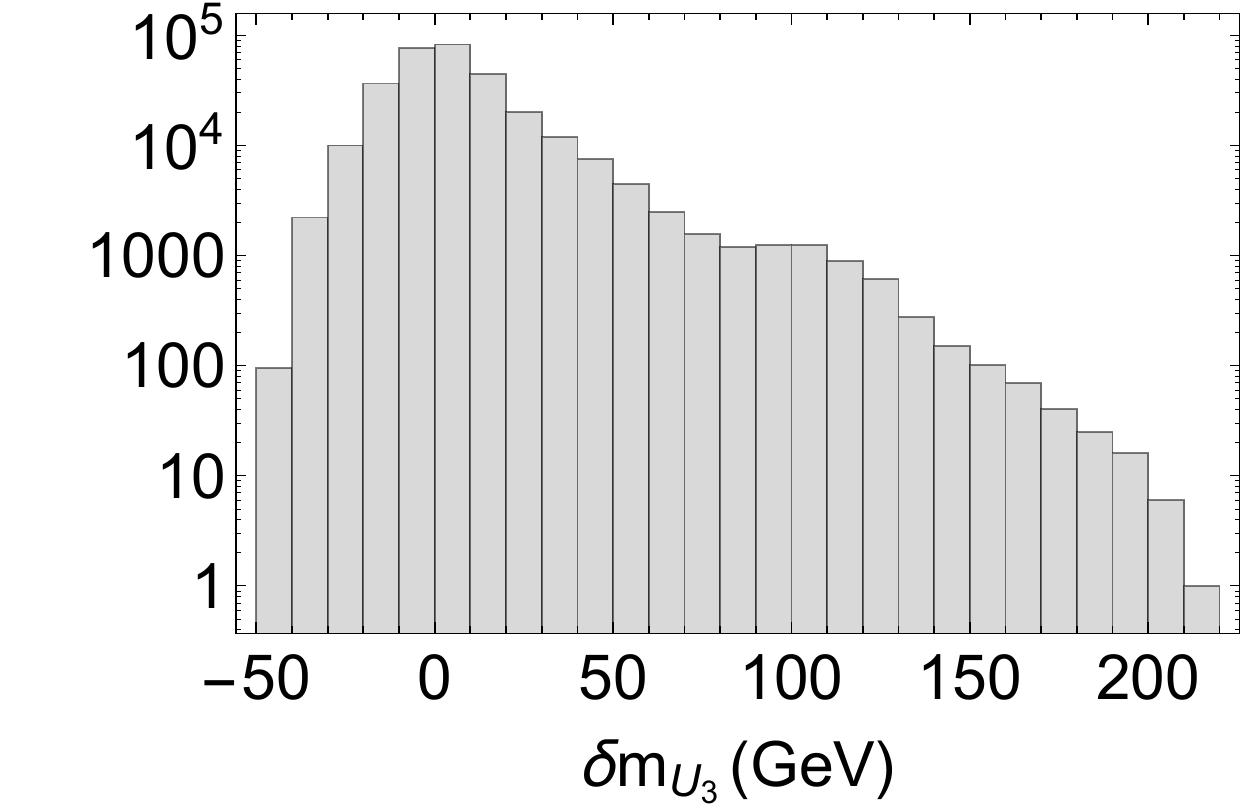}}
\caption{Histograms showing the changes in $m_{Q_3}$ and $m_{U_3}$ after running them up to the messenger scale using the transfer matrix and then back down to the weak scale using SoftSUSY, for every point in our GGM scan ($M_{mess}=10^7$, $10^{11}$, $10^{15}$ GeV).  \label{TMplot}}
\end{minipage}

\vspace{0.6cm}
\begin{minipage}{\textwidth}
\subfigure{\includegraphics[width=0.45\textwidth]{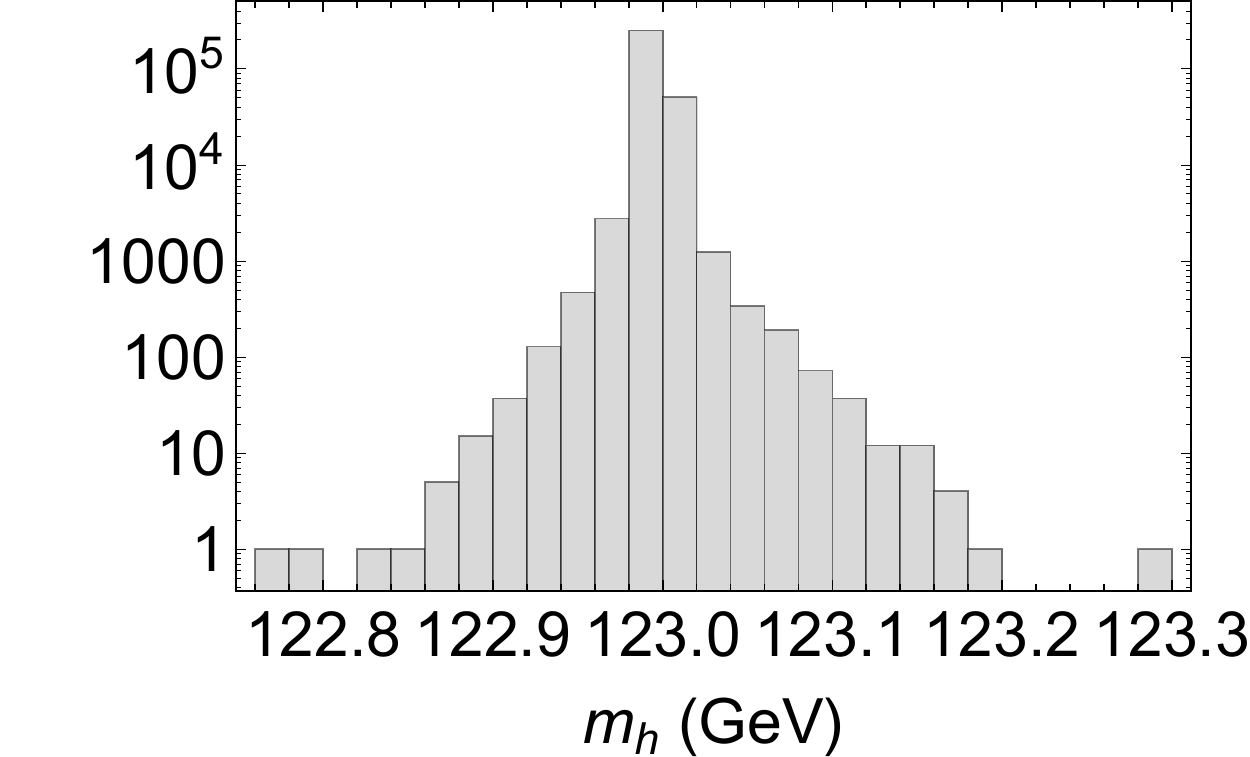}}\hfill
\subfigure{\raisebox{-3.2mm}{\includegraphics[width=0.45\textwidth]{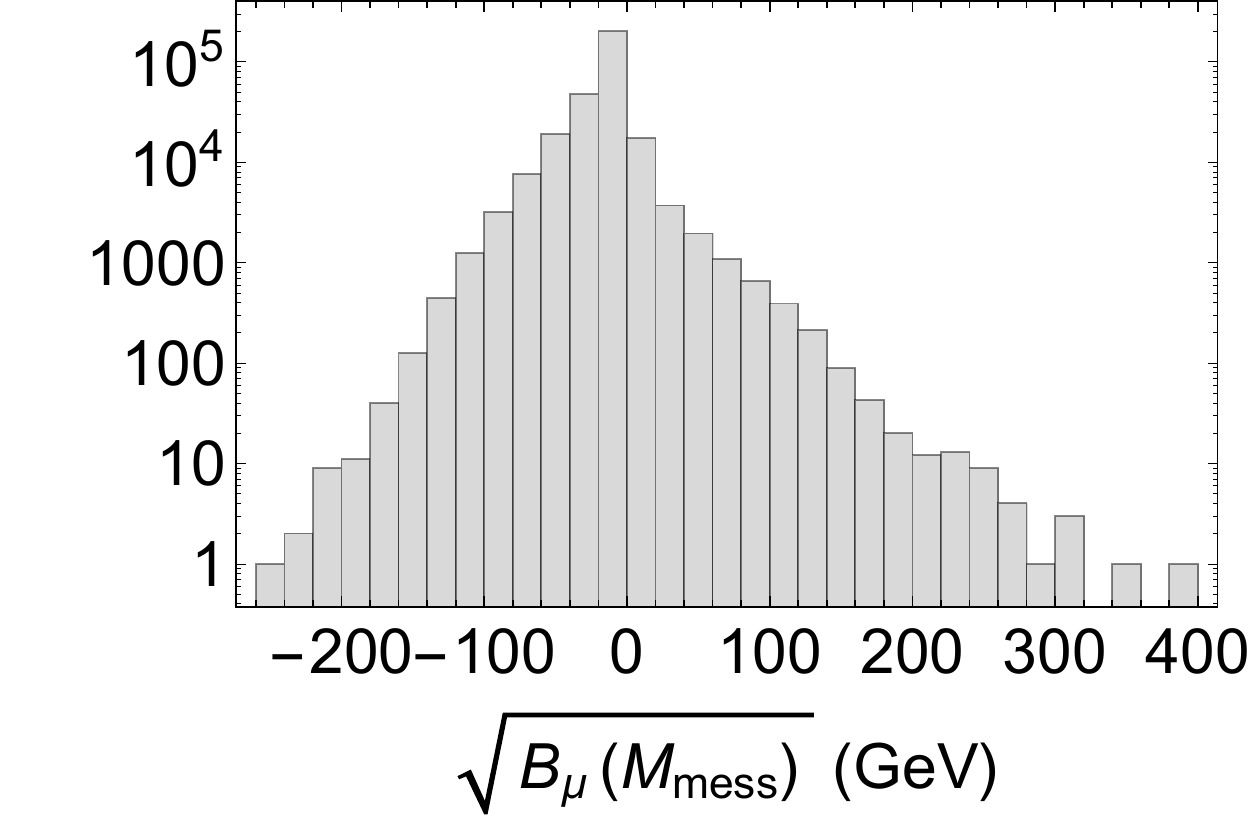}}}
\caption{Histograms of $m_h$ and $\sqrt{B_\mu({\rm UV})}$ across the entire GGM parameter scan. \label{convergenceplot}}
\end{minipage}

\end{figure}

 Shown in fig.~\ref{TMplot} are distributions of $\delta m_{Q_3}$ and $\delta m_{U_3}$ across our entire scan of GGM parameter space, which contains $\sim 3\times 10^5$ points. Here $\delta m_{Q_3}$ and  $\delta m_{U_3}$ are the change in $m_{Q_3}$ and $m_{U_3}$ after running them up to the messenger scale using the transfer matrix and then back down to the weak scale using SoftSUSY. We see that the accuracy of the transfer matrix is very good, generally differing by less than $\sim 50$~GeV, and never differing by more than $\sim 200$~GeV across the entire parameter space. These minor differences are due to effects not captured by the transfer matrix, primarily SoftSUSY's iterative determination of $M_S$, and IR threshold corrections to $g_3$ and $y_t$.

Shown in fig.~\ref{convergenceplot} are $m_h$ and $\sqrt{B_\mu({\rm UV})}$ for every point in our GGM parameter scan. We see that the convergence on $m_h$ is excellent, and the convergence on $\sqrt{B_\mu({\rm UV})}$ is decent (99\% of the points have $|\sqrt{B_\mu({\rm UV})}| < 400$~GeV). We also note in passing that the accuracy of the numerical scan is easily comparable to or larger than a naive estimate of the higher-loop $B_\mu(UV)$ expected from GGM. Given that the numerical scan was validated using the semi-analytic approach which assumed $B_\mu(UV)=0$, we do not expect that the small corrections to $B_\mu(UV)$ from GGM will make any difference to our conclusions.

\FloatBarrier

\section{Chargino/neutralino contribution to $m_h$}\label{appChargino/neutralino}

In this appendix we will delve into the threshold correction to $m_h$ from light winos and Higgsinos that greatly reduces the $A$-term required to achieve $m_h=123$~GeV in the neighborhood of $M_2=0$.

\begin{figure}[b!]\centering
\includegraphics[width=0.55\textwidth]{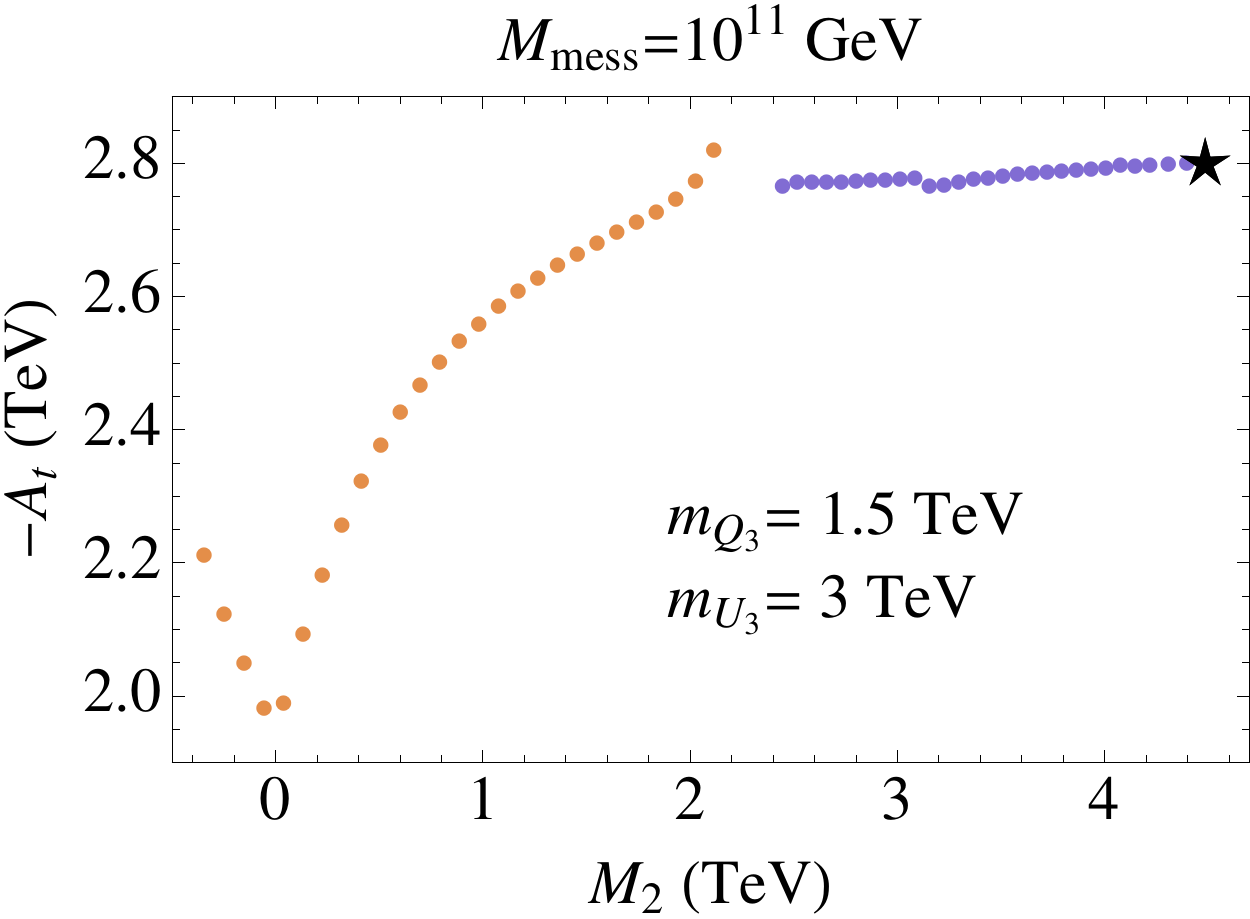}
\caption{Plot of $-A_t$ vs $M_2$ for the benchmark point indicated in the figure. Orange (blue) points correspond to $\mu>0$ ($\mu<0$). The black star corresponds to rightmost point in the $M_2$ interval which has the biggest $A_t$.}
  \label{At_Mmess11_mQ_1500_mU_3500}
\end{figure}

Shown in  fig.~\ref{At_Mmess11_mQ_1500_mU_3500} is a plot of $-A_t$  vs $M_2$ for a benchmark point extracted from our grid with $M_{mess}=10^{11}\text{ GeV}$ and $(m_{Q_3},m_{U_3})=(1.5,3.5)\text{ TeV}$. We see that $A_t$ varies by $\sim 40\%$  for $\mu>0$ (orange points), yet only varies by $\sim 1.5\%$ for $\mu<0$ (blue points). The variation for $\mu>0$ comes in the form of a sharp decrease in the magnitude of $A_t$ as we move from large $M_2$ to small $M_2$. This is characteristic of much of the parameter space, as we already saw in fig.~\ref{AtvsM2}.

\begin{figure}[p]\centering
\includegraphics[width=0.48\textwidth]{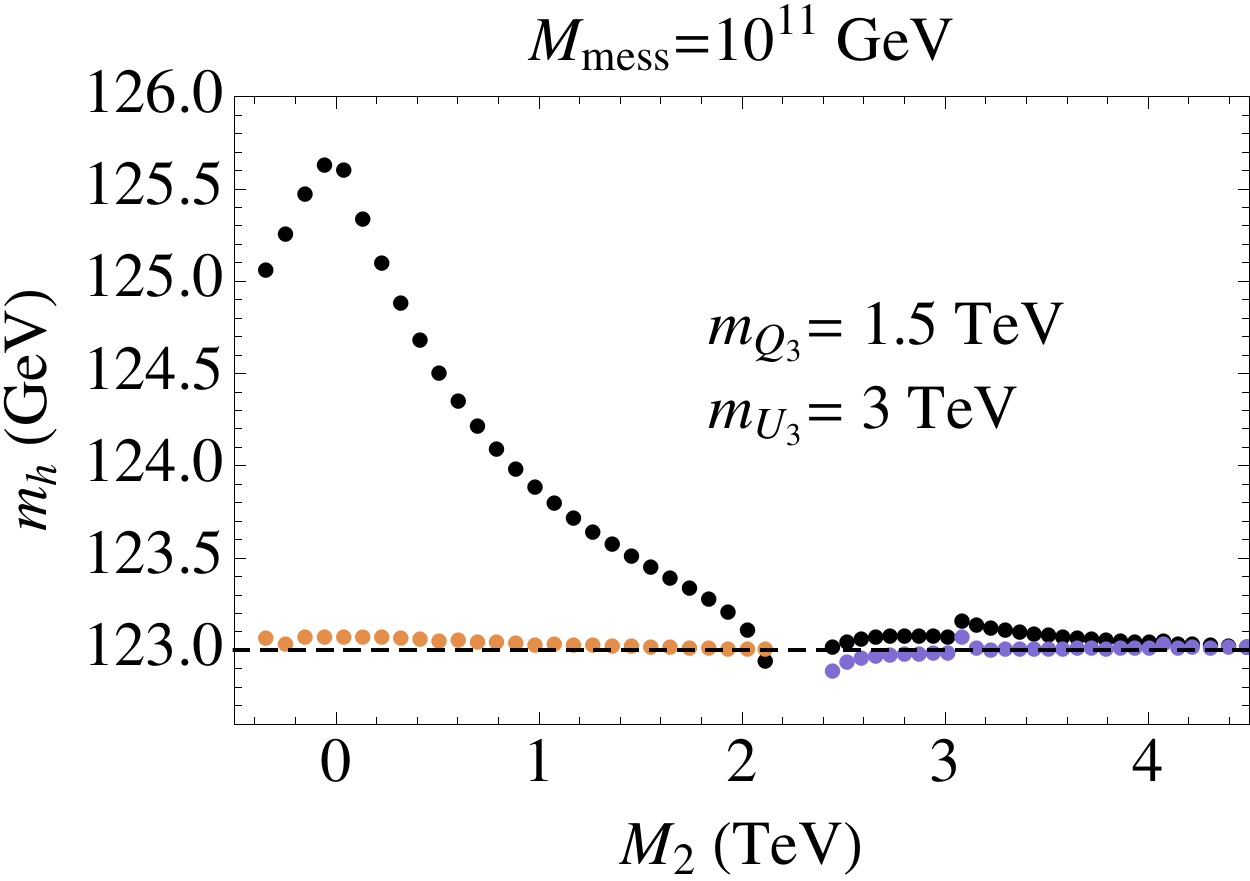}\quad  \includegraphics[width=0.48\textwidth]{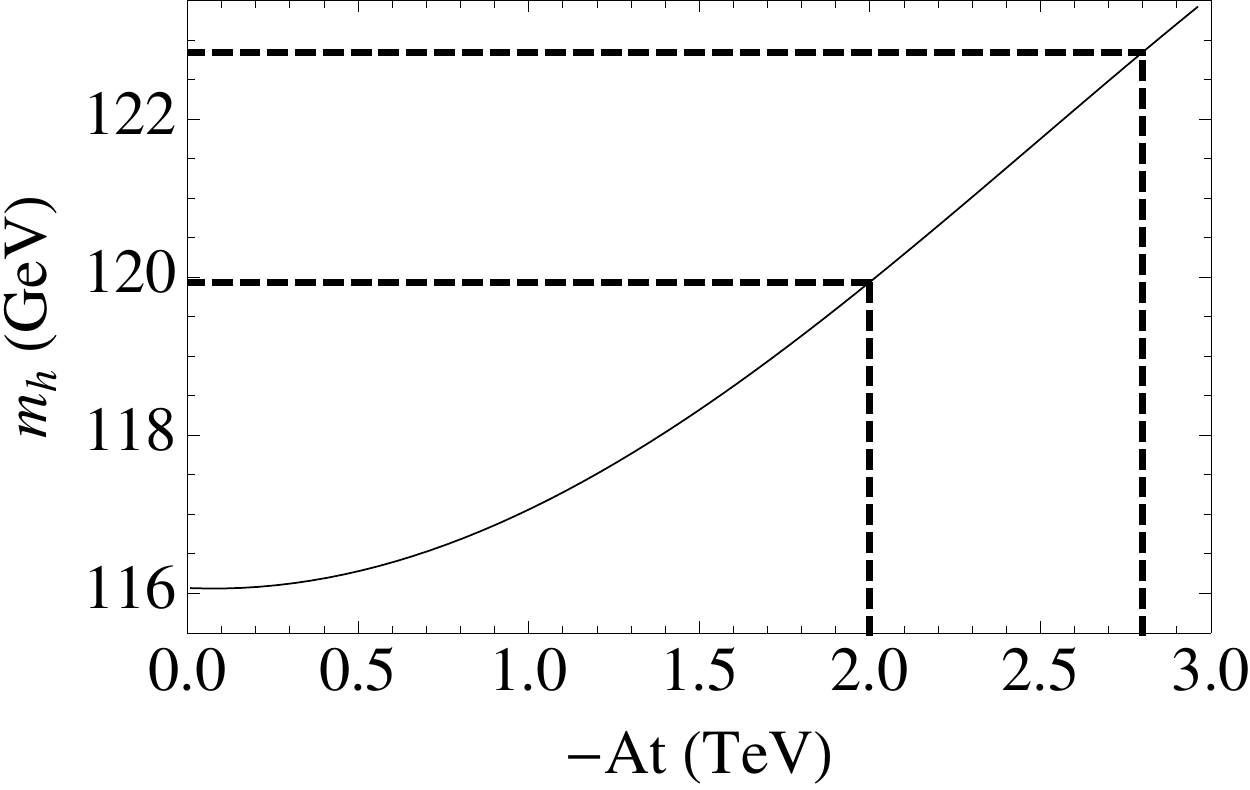}
\caption{{\bf Left:} Plot of $m_h$  vs $M_2$. Orange (blue) dots are grid points with $\mu>0$ ($\mu<0$). Black dots have $A_t=-2.8$~TeV as in the right-most point of the $M_2$ interval (the black star in fig.~\ref{At_Mmess11_mQ_1500_mU_3500}). {\bf Right:} Plot of $m_h$  vs $-A_t$. We fix all the other parameters as in the right-most point of the $M_2$ interval in fig.~\ref{At_Mmess11_mQ_1500_mU_3500}. The dashed lines indicate the enhancement of the Higgs mass induced by the drop of $M_2$ which can be extracted again from fig.~\ref{At_Mmess11_mQ_1500_mU_3500}.}

  \label{mh_vs_AtM2}
\end{figure}

Clearly, the decrease in magnitude of $A_t$ must be driven by a positive correction to the Higgs mass in the neighborhood of $M_2=0$, as shown in fig.~\ref{mh_vs_AtM2}. On the left, we have a plot of $m_h$ vs $M_2$ for the same benchmark of fig.~\ref{At_Mmess11_mQ_1500_mU_3500}. We see again the excellent convergence on $m_h=123$~GeV.  Meanwhile in black we plot the same points, but hold  $A_t$ fixed at $-2.8$~TeV, which corresponds to the right-most point on the $M_2$ interval (indicated with a black star in fig.~\ref{At_Mmess11_mQ_1500_mU_3500}). We see that as we move towards $M_2=0$, if we don't decrease the magnitude of $A_t$ to compensate for the effect at $M_2\approx0$, then the Higgs mass increases by as much as $\sim 2.5$~GeV.

Alternatively one can see the same effect on the right plot of fig.~\ref{mh_vs_AtM2}. Here we show $m_h$ vs $A_t$ with all other soft parameters fixed to the values corresponding to the right-most point on the $M_2$ interval of fig.~\ref{At_Mmess11_mQ_1500_mU_3500}. We see that from $A_t= -2.8$~TeV  to $A_t=-2$~TeV, $m_h$ decreases by 2.5~GeV. So we confirm that the origin of the variation in $A_t$ is a $\sim 2.5$~GeV enhancement to $m_h$ as we move to $M_2\to 0$.

\begin{figure}[p]\centering
\includegraphics[width=0.65\textwidth]{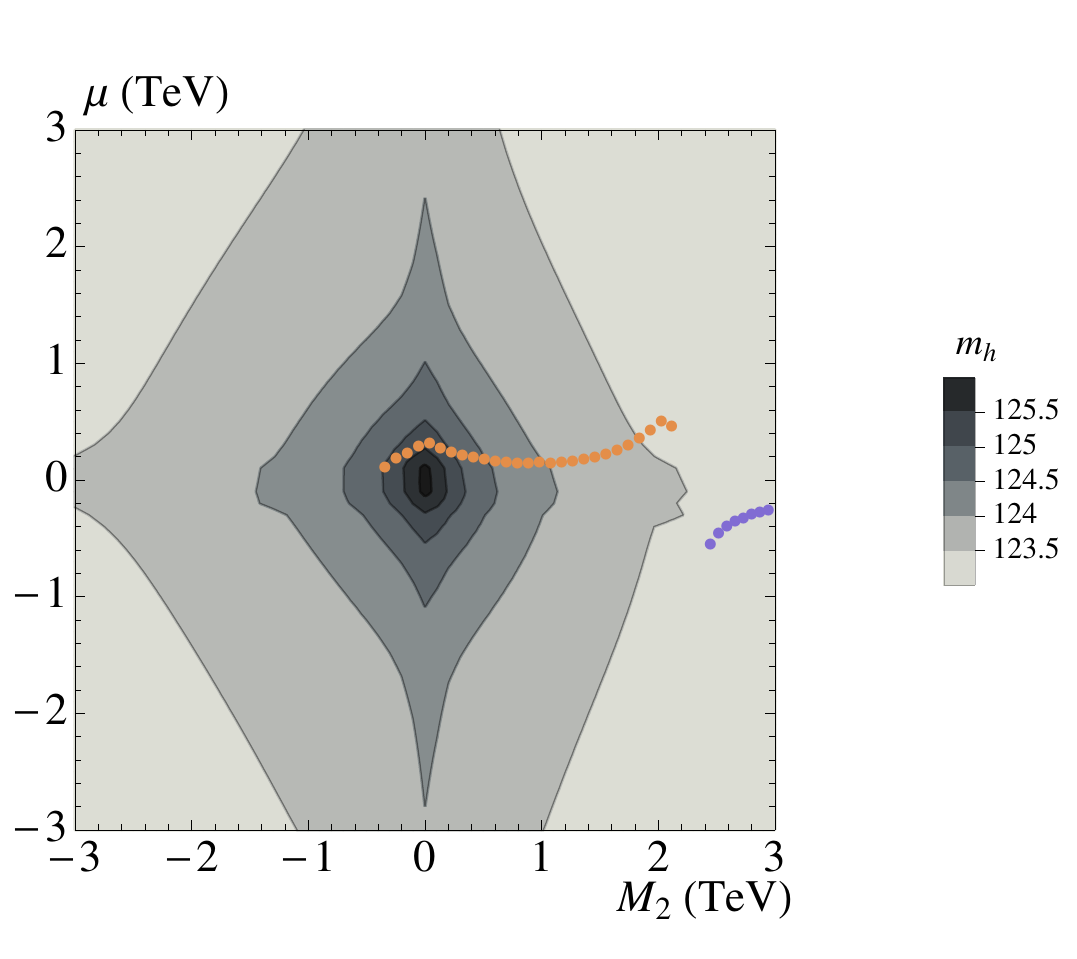}

\caption{Contour plot of $m_h$ in the ($M_2$, $\mu$) plane, with all other parameters fixed to those of the right-most point on the $M_2$ interval (the black star in fig.~\ref{At_Mmess11_mQ_1500_mU_3500}) and $X_t=A_t-y_t\mu\cot\beta$ kept fixed to $-2.8$ TeV. Overlaid in orange (blue) are the actual values of $M_2$ and $\mu$ for the benchmark point in fig.~\ref{At_Mmess11_mQ_1500_mU_3500} for $\mu>0$ ($\mu<0$). }
  \label{mh_vs_M2_mu}
\end{figure}

It remains to isolate the origin of the 2.5~GeV threshold correction to $m_h$. Shown in fig.~\ref{mh_vs_M2_mu} is a contour plot of $m_h$ vs $\mu$ and $M_2$, with all other parameters fixed to those of the right-most point on the $M_2$ interval. We see that the threshold correction is due to light charginos and neutralinos, and both light higgsinos and winos (i.e.\ both small $\mu$ and $M_2$) are required for the full effect. A similar effect was recently mentioned in the context of EFT calculations of  the Higgs mass in \cite{Vega:2015fna}. We have further verified that this is the origin of the threshold correction, by direct calculation of the relevant one-loop diagrams as done in \cite{Pierce:1996zz}.

We conclude that the sharp decrease in the magnitude of $A_t$ for $\mu>0$ is fully warranted. It corresponds to a few GeV threshold correction to the Higgs mass in the neighborhood of $M_2=0$ from light higgsinos and winos. It would be interesting to explore the implications of this threshold correction further in a more general context. Perhaps it could provide another motivation for EWKino searches at the LHC.

\FloatBarrier

\providecommand{\href}[2]{#2}\begingroup\raggedright\endgroup


\begin{thebibliography}{10}

\bibitem{Aad:2012tfa}
{\bf ATLAS Collaboration} Collaboration, G.~Aad et~al., {\it {Observation of a
  new particle in the search for the Standard Model Higgs boson with the ATLAS
  detector at the LHC}},  {\em Phys.Lett.} {\bf B716} (2012) 1--29,
  [\href{http://arxiv.org/abs/1207.7214}{{\tt arXiv:1207.7214}}].

\bibitem{Chatrchyan:2012ufa}
{\bf CMS Collaboration} Collaboration, S.~Chatrchyan et~al., {\it {Observation
  of a new boson at a mass of 125 GeV with the CMS experiment at the LHC}},
  {\em Phys.Lett.} {\bf B716} (2012) 30--61,
  [\href{http://arxiv.org/abs/1207.7235}{{\tt arXiv:1207.7235}}].

\bibitem{Hall:2011aa}
L.~J. Hall, D.~Pinner, and J.~T. Ruderman, {\it {A Natural SUSY Higgs Near 126
  GeV}},  {\em JHEP} {\bf 1204} (2012) 131,
  [\href{http://arxiv.org/abs/1112.2703}{{\tt arXiv:1112.2703}}].

\bibitem{Heinemeyer:2011aa}
S.~Heinemeyer, O.~Stal, and G.~Weiglein, {\it {Interpreting the LHC Higgs
  Search Results in the MSSM}},  {\em Phys.Lett.} {\bf B710} (2012) 201--206,
  [\href{http://arxiv.org/abs/1112.3026}{{\tt arXiv:1112.3026}}].

\bibitem{Arbey:2011ab}
A.~Arbey, M.~Battaglia, A.~Djouadi, F.~Mahmoudi, and J.~Quevillon, {\it
  {Implications of a 125 GeV Higgs for supersymmetric models}},  {\em
  Phys.Lett.} {\bf B708} (2012) 162--169,
  [\href{http://arxiv.org/abs/1112.3028}{{\tt arXiv:1112.3028}}].

\bibitem{Arbey:2011aa}
A.~Arbey, M.~Battaglia, and F.~Mahmoudi, {\it {Constraints on the MSSM from the
  Higgs Sector}},  {\em Eur.Phys.J.} {\bf C72} (2012) 1906,
  [\href{http://arxiv.org/abs/1112.3032}{{\tt arXiv:1112.3032}}].

\bibitem{Draper:2011aa}
P.~Draper, P.~Meade, M.~Reece, and D.~Shih, {\it {Implications of a 125 GeV
  Higgs for the MSSM and Low-Scale SUSY Breaking}},  {\em Phys.Rev.} {\bf D85}
  (2012) 095007, [\href{http://arxiv.org/abs/1112.3068}{{\tt
  arXiv:1112.3068}}].

\bibitem{Carena:2011aa}
M.~Carena, S.~Gori, N.~R. Shah, and C.~E. Wagner, {\it {A 125 GeV SM-like Higgs
  in the MSSM and the $\gamma \gamma$ rate}},  {\em JHEP} {\bf 1203} (2012)
  014, [\href{http://arxiv.org/abs/1112.3336}{{\tt arXiv:1112.3336}}].

\bibitem{Cao:2012fz}
J.~Cao, Z.~Heng, J.~M. Yang, Y.~Zhang, and J.~Zhu, {\it {A SM-like Higgs near
  125 GeV in low energy SUSY: a comparative study for MSSM and NMSSM}},  {\em
  JHEP} {\bf 1203} (2012) 086, [\href{http://arxiv.org/abs/1202.5821}{{\tt
  arXiv:1202.5821}}].

\bibitem{Christensen:2012ei}
N.~D. Christensen, T.~Han, and S.~Su, {\it {MSSM Higgs Bosons at The LHC}},
  {\em Phys.Rev.} {\bf D85} (2012) 115018,
  [\href{http://arxiv.org/abs/1203.3207}{{\tt arXiv:1203.3207}}].

\bibitem{Brummer:2012ns}
F.~Brummer, S.~Kraml, and S.~Kulkarni, {\it {Anatomy of maximal stop mixing in
  the MSSM}},  {\em JHEP} {\bf 1208} (2012) 089,
  [\href{http://arxiv.org/abs/1204.5977}{{\tt arXiv:1204.5977}}].

\bibitem{Evans:2011bea}
J.~L. Evans, M.~Ibe, and T.~T. Yanagida, {\it {Relatively Heavy Higgs Boson in
  More Generic Gauge Mediation}},  {\em Phys.Lett.} {\bf B705} (2011) 342--348,
  [\href{http://arxiv.org/abs/1107.3006}{{\tt arXiv:1107.3006}}].

\bibitem{Evans:2012hg}
J.~L. Evans, M.~Ibe, S.~Shirai, and T.~T. Yanagida, {\it {A 125GeV Higgs Boson
  and Muon g-2 in More Generic Gauge Mediation}},  {\em Phys.Rev.} {\bf D85}
  (2012) 095004, [\href{http://arxiv.org/abs/1201.2611}{{\tt
  arXiv:1201.2611}}].

\bibitem{Kang:2012ra}
Z.~Kang, T.~Li, T.~Liu, C.~Tong, and J.~M. Yang, {\it {A Heavy SM-like Higgs
  and a Light Stop from Yukawa-Deflected Gauge Mediation}},  {\em Phys.Rev.}
  {\bf D86} (2012) 095020, [\href{http://arxiv.org/abs/1203.2336}{{\tt
  arXiv:1203.2336}}].

\bibitem{Craig:2012xp}
N.~Craig, S.~Knapen, D.~Shih, and Y.~Zhao, {\it {A Complete Model of Low-Scale
  Gauge Mediation}},  {\em JHEP} {\bf 1303} (2013) 154,
  [\href{http://arxiv.org/abs/1206.4086}{{\tt arXiv:1206.4086}}].

\bibitem{Abdullah:2012tq}
M.~Abdullah, I.~Galon, Y.~Shadmi, and Y.~Shirman, {\it {Flavored Gauge
  Mediation, A Heavy Higgs, and Supersymmetric Alignment}},  {\em JHEP} {\bf
  1306} (2013) 057, [\href{http://arxiv.org/abs/1209.4904}{{\tt
  arXiv:1209.4904}}].

\bibitem{Kim:2012vz}
H.~D. Kim, D.~Y. Mo, and M.-S. Seo, {\it {Neutrino Assisted Gauge Mediation}},
  {\em Eur.Phys.J.} {\bf C73} (2013), no.~6 2449,
  [\href{http://arxiv.org/abs/1211.6479}{{\tt arXiv:1211.6479}}].

\bibitem{Byakti:2013ti}
P.~Byakti and T.~S. Ray, {\it {Burgeoning the Higgs mass to 125 GeV through
  messenger-matter interactions in GMSB models}},  {\em JHEP} {\bf 1305} (2013)
  055, [\href{http://arxiv.org/abs/1301.7605}{{\tt arXiv:1301.7605}}].

\bibitem{Craig:2013wga}
N.~Craig, S.~Knapen, and D.~Shih, {\it {General Messenger Higgs Mediation}},
  {\em JHEP} {\bf 1308} (2013) 118, [\href{http://arxiv.org/abs/1302.2642}{{\tt
  arXiv:1302.2642}}].

\bibitem{Evans:2013kxa}
J.~A. Evans and D.~Shih, {\it {Surveying Extended GMSB Models with
  $m$$_{h}$=125 GeV}},  {\em JHEP} {\bf 1308} (2013) 093,
  [\href{http://arxiv.org/abs/1303.0228}{{\tt arXiv:1303.0228}}].

\bibitem{Calibbi:2013mka}
L.~Calibbi, P.~Paradisi, and R.~Ziegler, {\it {Gauge Mediation beyond Minimal
  Flavor Violation}},  {\em JHEP} {\bf 1306} (2013) 052,
  [\href{http://arxiv.org/abs/1304.1453}{{\tt arXiv:1304.1453}}].

\bibitem{Jelinski:2013kta}
T.~Jelinski, {\it {On messengers couplings in extended GMSB models}},  {\em
  JHEP} {\bf 1309} (2013) 107, [\href{http://arxiv.org/abs/1305.6277}{{\tt
  arXiv:1305.6277}}].

\bibitem{Galon:2013jba}
I.~Galon, G.~Perez, and Y.~Shadmi, {\it {Non-Degenerate Squarks from Flavored
  Gauge Mediation}},  {\em JHEP} {\bf 1309} (2013) 117,
  [\href{http://arxiv.org/abs/1306.6631}{{\tt arXiv:1306.6631}}].

\bibitem{Fischler:2013tva}
W.~Fischler and W.~Tangarife, {\it {Vector-like Fields, Messenger Mixing and
  the Higgs mass in Gauge Mediation}},  {\em JHEP} {\bf 1405} (2014) 151,
  [\href{http://arxiv.org/abs/1310.6369}{{\tt arXiv:1310.6369}}].

\bibitem{Knapen:2013zla}
S.~Knapen and D.~Shih, {\it {Higgs Mediation with Strong Hidden Sector
  Dynamics}},  {\em JHEP} {\bf 1408} (2014) 136,
  [\href{http://arxiv.org/abs/1311.7107}{{\tt arXiv:1311.7107}}].

\bibitem{Ding:2013pya}
R.~Ding, T.~Li, F.~Staub, and B.~Zhu, {\it {Focus Point Supersymmetry in
  Extended Gauge Mediation}},  {\em JHEP} {\bf 1403} (2014) 130,
  [\href{http://arxiv.org/abs/1312.5407}{{\tt arXiv:1312.5407}}].

\bibitem{Calibbi:2014yha}
L.~Calibbi, P.~Paradisi, and R.~Ziegler, {\it {Lepton Flavor Violation in
  Flavored Gauge Mediation}},  {\em Eur.Phys.J.} {\bf C74} (2014), no.~12 3211,
  [\href{http://arxiv.org/abs/1408.0754}{{\tt arXiv:1408.0754}}].

\bibitem{Basirnia:2015vga}
A.~Basirnia, D.~Egana-Ugrinovic, S.~Knapen, and D.~Shih, {\it {125 GeV Higgs
  from Tree-Level $A$-terms}},  [\href{http://arxiv.org/abs/1501.00997}{{\tt
  arXiv:1501.00997}}].

\bibitem{Jelinski:2015gsa}
T.~Jelinski, {\it {$SO(10)$ inspired extended GMSB models}},
  [\href{http://arxiv.org/abs/1505.06722}{{\tt arXiv:1505.06722}}].

\bibitem{Jelinski:2015voa}
T.~Jelinski and J.~Gluza, {\it {Analytical two-loop soft mass terms of
  sfermions in Extended GMSB models}},
 [\href{http://arxiv.org/abs/1505.07443}{{\tt arXiv:1505.07443}}].

\bibitem{Giudice:1998bp}
G.~Giudice and R.~Rattazzi, {\it {Theories with gauge mediated supersymmetry
  breaking}},  {\em Phys.Rept.} {\bf 322} (1999) 419--499,
  [\href{http://arxiv.org/abs/hep-ph/9801271}{{\tt hep-ph/9801271}}].

\bibitem{Meade:2008wd}
P.~Meade, N.~Seiberg, and D.~Shih, {\it {General Gauge Mediation}},  {\em
  Prog.Theor.Phys.Suppl.} {\bf 177} (2009) 143--158,
  [\href{http://arxiv.org/abs/0801.3278}{{\tt arXiv:0801.3278}}].

\bibitem{Buican:2008ws}
M.~Buican, P.~Meade, N.~Seiberg, and D.~Shih, {\it {Exploring General Gauge
  Mediation}},  {\em JHEP} {\bf 0903} (2009) 016,
  [\href{http://arxiv.org/abs/0812.3668}{{\tt arXiv:0812.3668}}].

\bibitem{Komargodski:2008ax}
Z.~Komargodski and N.~Seiberg, {\it {mu and General Gauge Mediation}},  {\em
  JHEP} {\bf 0903} (2009) 072, [\href{http://arxiv.org/abs/0812.3900}{{\tt
  arXiv:0812.3900}}].

\bibitem{GGMcollider}
S.~Knapen, D.~Redigolo, and D.~Shih, {\it {to appear}}, .

\bibitem{Abel:2009ve}
S.~Abel, M.~J. Dolan, J.~Jaeckel, and V.~V. Khoze, {\it {Phenomenology of Pure
  General Gauge Mediation}},  {\em JHEP} {\bf 0912} (2009) 001,
  [\href{http://arxiv.org/abs/0910.2674}{{\tt arXiv:0910.2674}}].

\bibitem{Abel:2010vba}
S.~Abel, M.~J. Dolan, J.~Jaeckel, and V.~V. Khoze, {\it {Pure General Gauge
  Mediation for Early LHC Searches}},  {\em JHEP} {\bf 1012} (2010) 049,
  [\href{http://arxiv.org/abs/1009.1164}{{\tt arXiv:1009.1164}}].

\bibitem{Grajek:2013ola}
P.~Grajek, A.~Mariotti, and D.~Redigolo, {\it {Phenomenology of General Gauge
  Mediation in light of a 125 GeV Higgs}},  {\em JHEP} {\bf 1307} (2013) 109,
  [\href{http://arxiv.org/abs/1303.0870}{{\tt arXiv:1303.0870}}].

\bibitem{Rajaraman:2009ga}
A.~Rajaraman, Y.~Shirman, J.~Smidt, and F.~Yu, {\it {Parameter Space of General
  Gauge Mediation}},  {\em Phys.Lett.} {\bf B678} (2009) 367--372,
  [\href{http://arxiv.org/abs/0903.0668}{{\tt arXiv:0903.0668}}].

\bibitem{Carpenter:2008he}
L.~M. Carpenter, {\it {Surveying the Phenomenology of General Gauge
  Mediation}},  [\href{http://arxiv.org/abs/0812.2051}{{\tt arXiv:0812.2051}}].

\bibitem{Carena:2010gr}
M.~Carena, P.~Draper, N.~R. Shah, and C.~E. Wagner, {\it {Determining the
  Structure of Supersymmetry-Breaking with Renormalization Group Invariants}},
  {\em Phys.Rev.} {\bf D82} (2010) 075005,
  [\href{http://arxiv.org/abs/1006.4363}{{\tt arXiv:1006.4363}}].

\bibitem{Carena:2010wv}
M.~Carena, P.~Draper, N.~R. Shah, and C.~E. Wagner, {\it {SUSY-Breaking
  Parameters from RG Invariants at the LHC}},  {\em Phys.Rev.} {\bf D83} (2011)
  035014, [\href{http://arxiv.org/abs/1011.4958}{{\tt arXiv:1011.4958}}].

\bibitem{Allanach:2001kg}
B.~Allanach, {\it {SOFTSUSY: a program for calculating supersymmetric
  spectra}},  {\em Comput.Phys.Commun.} {\bf 143} (2002) 305--331,
  [\href{http://arxiv.org/abs/hep-ph/0104145}{{\tt hep-ph/0104145}}].

\bibitem{Dermisek:2006ey}
R.~Dermisek and H.~D. Kim, {\it {Radiatively generated maximal mixing scenario
  for the Higgs mass and the least fine tuned minimal supersymmetric standard
  model}},  {\em Phys.Rev.Lett.} {\bf 96} (2006) 211803,
  [\href{http://arxiv.org/abs/hep-ph/0601036}{{\tt hep-ph/0601036}}].

\bibitem{Calibbi:2014pza}
L.~Calibbi, A.~Mariotti, C.~Petersson, and D.~Redigolo, {\it {Selectron NLSP in
  Gauge Mediation}},  {\em JHEP} {\bf 1409} (2014) 133,
  [\href{http://arxiv.org/abs/1405.4859}{{\tt arXiv:1405.4859}}].

\bibitem{Pierce:1996zz}
D.~M. Pierce, J.~A. Bagger, K.~T. Matchev, and R.-j. Zhang, {\it {Precision
  corrections in the minimal supersymmetric standard model}},  {\em Nucl.Phys.}
  {\bf B491} (1997) 3--67, [\href{http://arxiv.org/abs/hep-ph/9606211}{{\tt
  hep-ph/9606211}}].

\bibitem{Allanach:2004rh}
B.~Allanach, A.~Djouadi, J.~Kneur, W.~Porod, and P.~Slavich, {\it {Precise
  determination of the neutral Higgs boson masses in the MSSM}},  {\em JHEP}
  {\bf 0409} (2004) 044, [\href{http://arxiv.org/abs/hep-ph/0406166}{{\tt
  hep-ph/0406166}}].

\bibitem{Liu:2015bma}
Z.~Liu and B.~Tweedie, {\it {The Fate of Long-Lived Superparticles with
  Hadronic Decays after LHC Run 1}},  {\em JHEP} {\bf 1506} (2015) 042,
  [\href{http://arxiv.org/abs/1503.05923}{{\tt arXiv:1503.05923}}].

\bibitem{Dimopoulos:1996ig}
S.~Dimopoulos and G.~Giudice, {\it {Multimessenger theories of gauge mediated
  supersymmetry breaking}},  {\em Phys.Lett.} {\bf B393} (1997) 72--78,
  [\href{http://arxiv.org/abs/hep-ph/9609344}{{\tt hep-ph/9609344}}].

\bibitem{Argurio:2012qt}
R.~Argurio and D.~Redigolo, {\it {Tame D-tadpoles in gauge mediation}},  {\em
  JHEP} {\bf 1301} (2013) 075, [\href{http://arxiv.org/abs/1206.7037}{{\tt
  arXiv:1206.7037}}].

\bibitem{Backovic:2015rwa}
M.~Backovic, A.~Mariotti, and M.~Spannowsky, {\it {Signs of Tops from Highly
  Mixed Stops}},  [\href{http://arxiv.org/abs/1504.00927}{{\tt
  arXiv:1504.00927}}].

\bibitem{Riotto:1995am}
A.~Riotto and E.~Roulet, {\it {Vacuum decay along supersymmetric flat
  directions}},  {\em Phys.Lett.} {\bf B377} (1996) 60--66,
  [\href{http://arxiv.org/abs/hep-ph/9512401}{{\tt hep-ph/9512401}}].

\bibitem{Vega:2015fna}
J.~P. Vega and G.~Villadoro, {\it {SusyHD: Higgs mass Determination in
  Supersymmetry}},  [\href{http://arxiv.org/abs/1504.05200}{{\tt
  arXiv:1504.05200}}].

\bibitem{Espinosa:2001mm}
J.~Espinosa and I.~Navarro, {\it {Radiative corrections to the Higgs boson mass
  for a hierarchical stop spectrum}},  {\em Nucl.Phys.} {\bf B615} (2001)
  82--116, [\href{http://arxiv.org/abs/hep-ph/0104047}{{\tt hep-ph/0104047}}].

\end{thebibliography}
\end{document}